\documentclass[aps,prd,onecolumn,eqsecnum,amsmath,nofootinbib,preprintnumbers]{revtex4}%

\usepackage{color,graphicx, float, subfigure, xcolor}
\usepackage{amsfonts,amssymb, theorem,mathrsfs, times}
\usepackage{bm}
\usepackage{multirow}
\usepackage{mathtools}
\usepackage{amsfonts,amssymb,theorem,mathrsfs}
\usepackage{dsfont}
\usepackage{setspace}
\usepackage{amsmath} 
\usepackage[colorlinks,linkcolor=blue,anchorcolor=blue,citecolor=blue]{hyperref}   
\usepackage{ulem}   
\usepackage[english]{babel}

\usepackage{epstopdf}
\usepackage{booktabs} 
\usepackage{tabularx} 
\usepackage{makecell}

{\theorembodyfont{\upshape}
	}
{\theorembodyfont{\upshape}
	}
{\theorembodyfont{\upshape}
	}
{\theorembodyfont{\upshape}
	}
{\theorembodyfont{\upshape}
	}
{\theorembodyfont{\upshape}
	}
\addtocounter{MaxMatrixCols}{10}
\newcommand{\dalm}{\kern1pt\vbox{\hrule height 0.9pt\hbox{\vrule width 0.9pt\hskip 2.5pt\vbox{\vskip 5.5pt}\hskip 3pt\vrule width 0.3pt}\hrule height 0.3pt}\kern1pt}

\begin{document}
\thispagestyle{empty}
\title{The quasinormal modes of the rotating quantum corrected black holes}
	
%



\author{Jia-Ning Chen$^{a\, ,b\, ,c}$\footnote{e-mail address: chenjianing22@mails.ucas.ac.cn}}   


\author{Zong-Kuan Guo$^{b\, ,a\, ,c}$\footnote{e-mail address: guozk@itp.ac.cn}}

\author{Liang-Bi Wu$^{a\, ,c}$\footnote{e-mail address: liangbi@mail.ustc.edu.cn (corresponding author)}}


	
\affiliation{${}^a$School of Fundamental Physics and Mathematical Sciences, Hangzhou Institute for Advanced Study, UCAS, Hangzhou 310024, China}

\affiliation{${}^b$Institute of Theoretical Physics, Chinese Academy of Sciences, Beijing 100190, China}

\affiliation{${}^c$University of Chinese Academy of Sciences, Beijing 100049, China}




\date{\today}
	
\begin{abstract}
The quasinormal modes (QNMs) of a rotating quantum corrected black hole (RQCBH) are studied by employing the hyperboloidal framework for the scalar perturbation. This framework is used to cast the QNMs spectra problem into a two-dimensional eigenvalue problem, then the spectra are calculated by imposing the two-dimensional pseudo-spectral method. Based on the resulting scalar spectra, a parameter estimation pipeline for this RQCBH model with gravitational wave data is constructed by using \texttt{pyRing} in the ringdown phase. We use informative priors in our inference that incorporates the mass and spin distributions predicted by the inspiral-merger phase as the prior distributions for the ringdown analysis. Notably, since the waveform model beyond Kerr black hole in $\texttt{pyRing}$ is designed for the tensor perturbation, the inferred posterior distributions should be interpreted as a methodological investigation rather than as physical constraints from observations. The methodological results show that the use of informative priors consistently yields a tighter posterior on the quantum correction parameter compared to analyses without such priors, and the spin inferred from the RQCBH model begins to be significant and differs from that of the Kerr model. This opens a promising avenue for testing quantum-gravity-induced deviations using gravitational-wave spectroscopy.




\end{abstract}

\maketitle
\section{Indroduction}
Although general relativity (GR) is a well-established theory within the classical domain and accounts for a wide range of gravitational phenomena, it remains unable to address the issue of spacetime singularities. Penrose's singularity theorem~\cite{Penrose:1964wq,Hawking:1970zqf} states that within the classical GR framework, the occurrence of singularities is inevitable, which leads to the breakdown of physical laws. Over the years, multiple strategies have been pursued to overcome the singularity issue, with the widespread belief that a quantum gravity theory could potentially provide the answer. Loop Quantum Gravity (LQG) stands out as one such proposed quantum gravity theory, defined by its background-independent nature and its non-perturbative formulation~\cite{Ashtekar:2004eh,Han:2005km}. Loop Quantum Cosmology (LQC), in particular, has yielded resolutions to the Big Bang singularity through the theoretical framework~\cite{Ashtekar:2003hd}. A recent investigation by Lewandowski \textit{et al.} focused on the gravitational collapse of a dust ball, employing a LQC framework to incorporate LQG effects~\cite{Lewandowski:2022zce}. This work provides key insights into the pivotal implications of quantum gravity for the behavior of collapsing structures. This work has inspired extensive follow-up studies across multiple aspects, including black hole shadows and images~\cite{Yang:2022btw,Zhang:2023okw,Ye:2023qks,Zhao:2024elr,You:2024jeu,Luo:2024nul,Ali:2024ssf,Raza:2025pgs,Ahmed:2025qor,Vachher:2024ait}, the instability of the inner horizon~\cite{Cao:2023aco}, thermodynamic properties~\cite{Wang:2024jtp,You:2024iwn,Tan:2024usd,Zhang:2024hyr,Tan:2025zsl,Ahmed:2025qor}, charged and higher-dimensional quantum Oppenheimer-Snyder models~\cite{Mazharimousavi:2025lld,Shi:2024vki}, the motion of massive particles~\cite{Yang:2024lmj,Alimova:2025izs,Yang:2025esa}, and the quasinormal modes (QNMs) or graybody factors~\cite{Yang:2022btw,Gingrich:2023fxu,Shao:2023qlt,Gong:2023ghh,Cao:2024oud,Skvortsova:2024atk,Luo:2024dxl,Ahmed:2025qor,Lv:2025nww,Dong:2024hod}. One can see other works about the quantum corrected black hole models for more inspiration~\cite{Bonanno:2025dry, Konoplya:2023ahd, Konoplya:2019xmn, Battista:2023iyu,Wang:2025fmz,Zi:2024jla}.

In this work, we first calculate the scalar QNMs of the rotating quantum corrected black hole (RQCBH) using the two-dimensional pseudo-spectral method, where the rotation version is obtained by the Newman-Janis generating method~\cite{Azreg-Ainou:2014aqa,Azreg-Ainou:2014pra} in a hyperboloidal framework. The hyperboloidal method transforms the perturbation equation into a hyperbolic partial differential equations~\cite{PanossoMacedo:2019npm}. After separating the time part, the QNMs spectra $\omega$ can be solved as a two-dimensional eigenvalue problem. There have been many studies on two-dimensional spectra problems of the Kerr-type black hole~\cite{Blazquez-Salcedo:2023hwg,Chung:2023wkd,Khoo:2024yeh,Chung:2024ira,Chung:2024vaf,Blazquez-Salcedo:2024oek,Blazquez-Salcedo:2024dur,Ripley:2022ypi,Cai:2025irl,Xiong:2024urw,Assaad:2025nbv}.

Furthermore, using observational data to estimate the range of these quantum correction parameters coming from the modified gravity theory is also an important issue, hoping to search the hidden codes that may transcend Einstein's theory. For such a quantum corrected parameter in~\cite{Lewandowski:2022zce}, there are already some references estimating this parameter. For example, using the Event Horizon Telescope (EHT) results, one can obtain the constraints on the parameters of such black hole~\cite{Ali:2024ssf,Vachher:2024ait,Raza:2025pgs,Zhao:2024elr}. In addition, constraints on this quantum corrected black hole with eccentric extreme mass-ratio inspirals have been derived in~\cite{Yang:2024lmj,Yang:2025esa}. However, there has been no study using QNMs spectra to constrain the quantum-corrected parameter. The ringdown phase serves as a crucial bridge between black hole perturbation theory~(resulting in the QNMs spectra) and time-domain gravitational-wave~(GW) observations. From GW150914~\cite{LIGOScientific:2016aoc, LIGOScientific:2016vlm} to GW250114~\cite{KAGRA:2025oiz}, with more than three hundred GW events accumulated, the understanding of black hole properties and of testing gravity theories has advanced rapidly over the past decade. During the latter half of this decade, the development of two key \textit{Python} packages, \texttt{pyRing}~\cite{Carullo:2019flw,Isi:2019aib,LIGOScientific:2020tif,pyRing} and \texttt{ringdown}~\cite{Isi:2019aib,Isi:2021iql},  has greatly accelerated the use of black hole data to probe various black hole properties. Both of the \textit{Python} packages are designed to perform Bayesian inference of black hole properties with gravitational QNMs and GW ringdown data. Although they exhibit some differences in the details of processing data pipelines~\cite{KAGRA:2025oiz}, their overall results remain consistent. In this work, we adopt scalar QNMs spectra in combination with GW data to perform parameter inference for the RQCBH using \texttt{pyRing}. Notablely, the mismatch in the type of QNMs spectra gives rise to the fact that the results should be interpreted as a methodological investigation rather
than as physical constraints on parameters from observations. While, as an initial investigation, this method also provides valuable insights for RQCBH and lays the groundwork for future analyses.

Additionally, we employ informative priors for ringdown inference. In principle, for a coalescence process of the binary black hole system, the pre-merger phase can provide informative priors for the post-merger stage. Along this direction, Refs. \cite{Pacilio:2024tdl, MaganaZertuche:2024ajz} investigated the impact of the inspiral-merger stage on the amplitudes and phases of the ringdown stage. Subsequently, the influence of the binary black hole parameters on the overall parameters of the ringdown stage has been studied using the gating-inpainting technique~\cite{Wang:2026rev}. We introduce the mass and spin distributions predicted from the inspiral-merger
phase as the informative prior distributions for the ringdown analysis. This choice improves the accuracy
of the ringdown modeling and enhances the ability to constrain the quantum corrected parameter.

This paper is organized as follows. In Sec. \ref{RQCBH}, we present the rotating quantum corrected black hole solution and give its basic property. In Sec. \ref{QNM_spectra}, the QNMs spectra of such RQCBH are given for the Klein-Gordon equation. In Sec. \ref{Bayesian_analysis}, we use the results of the QNMs spectra and do the Bayesian analysis and parameter estimation from several events. Conclusions and broader discussions are provided in Sec. \ref{Conclusions_and_discussion}. In Appendix \ref{height_functions}, the height function for the RQCBH is given, where the height function will be used to get the QNMs spectra from the hyperboloidal framework~\cite{PanossoMacedo:2019npm}. Furthermore, Appendix \ref{expressions_L1_and_L2} provides the explicit form of the operator $L$ (see the main text) used in the two-dimensional pseudo-spectral method. In Appendix \ref{numerical_accuracy_test}, we give numerical accuracy tests for the obtained results. Throughout this work, we use the mostly plus metric signature and adopt geometric units with $c=G=1$.

\section{Rotating quantum corrected black holes}\label{RQCBH}
In this section, we give a brief review on the so-called rotating quantum corrected black holes (RQCBH). We derive the RQCBH by using the updated Newman-Janis (NJ) generating method~\cite{Azreg-Ainou:2014aqa,Azreg-Ainou:2014pra}, which has been successful in generating the rotating solutions in Boyer-Lindquist (B-L) coordinates from spherically symmetric static solutions. For such a rotating black hole, the corresponding spherically symmetric static solution is derived from~\cite{Lewandowski:2022zce}, where its external metric reads
\begin{eqnarray}\label{spherically_symmetric_static_metric}
	\mathrm{d}s^2=-f(r)\mathrm{d}t^2+\frac{\mathrm{d}r^2}{f(r)}+r^2(\mathrm{d}\theta^2+\sin^2\theta\mathrm{d}\phi^2)\, ,
\end{eqnarray}
where the metric function $f(r)$ reads
\begin{eqnarray}\label{metric_function}
	f(r)=1-\frac{2M}{r}+\frac{\alpha M^2}{r^4}\, ,
\end{eqnarray}
with the parameter $\alpha=16\sqrt{3}\pi\gamma^3l^2_p$ , $l_p=\sqrt{\hbar}$ denoting the Planck length, $\gamma$ being the Immirzi parameter, and $M$ standing for the mass of the black hole. In the B-L coordinates $\{t,r,\theta,\varphi\}$, the corresponding RQCBH metric can be written as~\cite{Vachher:2024ait,Ali:2024ssf,Raza:2025pgs} 
\begin{eqnarray}\label{BL_coordinate}
	\mathrm{d}s^2&=&-\Big(\frac{\Delta-a^2\sin^2\theta}{\Sigma}\Big)\mathrm{d}t^2-2a\sin^2\theta\Big(1-\frac{\Delta-a^2\sin^2\theta}{\Sigma}\Big)\mathrm{d}t\mathrm{d}\varphi\nonumber\\
	&&+\sin^2\theta\Big[\Sigma+a^2\sin^2\theta\Big(2-\frac{\Delta-a^2\sin^2\theta}{\Sigma}\Big)\Big]\mathrm{d}\varphi^2+\frac{\Sigma}{\Delta}\mathrm{d}r^2+\Sigma\mathrm{d}\theta^2\, ,
\end{eqnarray}
where
\begin{eqnarray}\label{Delta_and_Sigma}
	\Delta=r^2+a^2-2Mr+\frac{\alpha M^2}{r^2}\, ,\quad \Sigma=r^2+a^2\cos^2\theta\, ,
\end{eqnarray}
and $a$ is the angular momentum. The above rotating metric provides important insights into the behavior of rotating black holes in the presence of quantum eﬀects. It is important to note that in the limit $a\to0$, RQCBH reduces to a spherical quantum corrected black hole~\cite{Lewandowski:2022zce} and in the limit $\alpha\to0$, RQCBH reduces to the Kerr spacetime. 

\begin{figure}[htbp]
	\centering
	\includegraphics[width=0.6\textwidth]{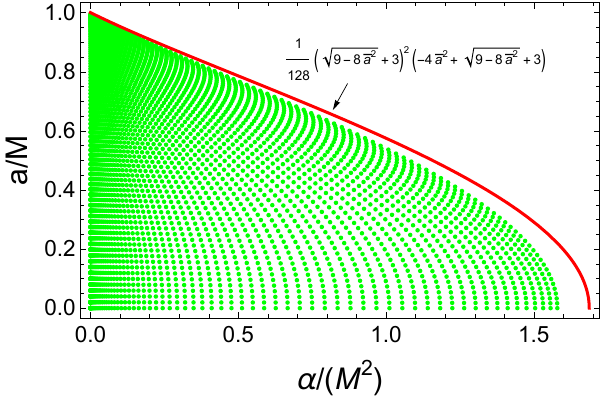}
	\caption{Parameter space $(\alpha/M^2,a/M)$ for RQCBH. The red solid line corresponds to the extremal black holes with degenerate horizons. The sample green points which are used to obtain the QNMs spectra in the parameter space $(\alpha/M^2,a/M)$ corresponding to the points $(q,\kappa)$ in the $q-\kappa$ plane (see Fig. \ref{parameter_region}), in which these points are confined to the set $\{(q,\kappa)|0\le\kappa\le q\, ,q\le0.9\}$.}
	\label{Samples_alpha_a}
\end{figure}

Depending on the values of $\alpha$ and $a$, there may be up to two positive roots of $\Delta$, and two horizons of RQCBH will be found. It's not difficult to notice that the quantum correction parameter $\alpha$ must be within the range $\alpha/M^2\le 27/16$ for the emergence of two horizons. In Fig. \ref{Samples_alpha_a}, we show the parameter space for RQCBH. For the parameters below the red solid line, the black hole has two event horizons. For the parameters above the red solid line, there is no horizon in spacetime. The parameters on the red solid line indicate that the black hole is extreme. For these green points inside, further explanation will be provided in Sec. \ref{QNM_spectra}. The red line is the boundary between black hole and non-horizon spacetime~\cite{Vachher:2024ait,Ali:2024ssf}. It is not difficult to find that its expression is
\begin{eqnarray} \label{eq: constrain on alpha and a}
    \bar{\alpha}(\bar{a})=\frac{1}{128} \Big(\sqrt{9-8 \bar{a}^2}+3\Big)^2 \Big(-4 \bar{a}^2+\sqrt{9-8 \bar{a}^2}+3\Big)\, ,\quad \bar{\alpha}\equiv\frac{\alpha}{M^2}\, ,\quad \bar{a}\equiv \frac{a}{M}\, .
\end{eqnarray}
Here, two parameter $\bar{\alpha}$ and $\bar{a}$ are both dimensionless.

The physical properties of the RQCBH had been discussed in Ref. \cite{Ali:2024ssf}, including the horizon structure and ergoregion, as well as the frame-dragging effect. The ergoregion $r_{+} < r < r_{\mathrm{SLS}}$~(Static Limit Surface) of RQCBH get larger with an increase in qunatume correction parameter $\alpha$ for fixed values of the parameter $a$, hence are larger than those for the Kerr black hole $(\alpha=0)$, it directly affects the negative energy orbits within the ergoregion, increasing the potential efficiency of energy extraction. The frame-dragging effect, characterized by the metric component $g_{t\phi}$, is also modified by the presence of $\alpha$. As an observer approaches the event horizon, it will ultimately co-rotate with RQCBH, whose final angular velocity is $\Omega_{\mathrm{RQCBH}} = {2 a r_{+}\Big(M-\frac{\alpha M^2}{2 r_{+}^3}\Big)}/{(r_{+}^2+a^2)^2}$. This velocity will reduce to the Kerr black hole angular velocity $\Omega_{\mathrm{Kerr}}= a/(r_{+}^{2} + a^{2})$ in the limit $\alpha \to 0$. Beyond these discussions, we also provide the surface gravity of the RQCBH, which takes the following form~\cite{Kumar:2020owy}
\begin{eqnarray}
                \kappa = \frac{\Delta^{\prime}(r)}{2(r^{2} + a^{2})} \bigg|_{r=r_{+}} = \frac{1}{r_{+}^{2} + a^2} \Big( r_{+} - M -\frac{\alpha M^{2}}{r_{+}^{3}}  \Big)  \, .
\end{eqnarray}
In the limit $\alpha \to 0$, the surface gravity of RQCBH reduces to $\kappa = \frac{r_{+}-M}{r_{+}^{2}+a^2}$ that corresponds to the surface gravity of the Kerr black hole. These effects are important for understanding the properties of the RQCBH as well as the related high-energy astrophysical phenomena.

On the other hand, the rotating solution corresponding to a spherical metric can also be constructed using the gravitational decoupling~(GD) approach~\cite{Contreras:2021yxe}. The metric function $f(r)$ of QCBH in the spherically symmetric metric can be written as $f(r) = 1 - \frac{2 \tilde{m}(r)}{r}$ with mass function $\tilde{m}(r) = M - \frac{\alpha M^2}{2 r^3}$. Then, substituting this mass function into the Gurses-Gursey metric~\cite{Gurses:1975vu} can obtain the corresponding rotating solution when the critical condition $\tilde{a}=a=a_{s}$ is satisfied. The solution using the GD approach is the same as the result constructed by the NJ algorithm. At the same time, the additional hair induced by $\alpha$ does not break the flatness of spacetime.

\section{QNMs spectra of Rotating quantum corrected black holes}\label{QNM_spectra}
We now proceed to an analysis of linearized perturbations of the RQCBH. Our focus will be on the QNMs spectra, for use in modelling the ringdown phase of the RQCBH. Given that this is a quantum effective solution, it is difficult to calculate general perturbations, so we study a test massless scalar field $\Phi$, which is an important simplifying approximation. This assumption allows us to focus on the Klein-Gordon equation for a massless scalar field given by
\begin{eqnarray}\label{KG_equation}
	\square \Phi=0\, .
\end{eqnarray}
For similar thought, one can refer to~\cite{Taylor:2024duw} in which a magnetic field is considered. The Klein-Gordon equation will be simplified in the B-L coordinates (\ref{BL_coordinate}) into
\begin{eqnarray}\label{KG_BL}
	&&\Big[\frac{(r^2+a^2)^2}{\Delta}-a^2\sin^2\theta\Big]\frac{\partial^2\Phi}{\partial t^2}+\frac{2aM\Big(2r-\frac{\alpha M}{r^2}\Big)}{\Delta}\frac{\partial^2\Phi}{\partial t\partial\varphi}\nonumber\\
	&&+\Big(\frac{a^2}{\Delta}-\frac{1}{\sin^2\theta}\Big)\frac{\partial^2\Phi}{\partial\varphi^2}-\frac{\partial}{\partial r}\Big(\Delta\frac{\partial\Phi}{\partial r}\Big)-\frac{1}{\sin\theta}\frac{\partial}{\partial \theta}\Big(\sin\theta\frac{\partial\Phi}{\partial\theta}\Big)=0\, .
\end{eqnarray}

It is known that there are many effective methods for getting the QNMs spectra, such as the continued fraction technique~\cite{Leaver:1985ax}, asymptotic iteration method~\cite{Cho:2009cj,Cho:2011sf}, Wentzel-Kramers-Brillouin (WKB) approximation~\cite{Konoplya:2003ii,Konoplya:2019hlu} and the pseudo-spectral method~\cite{Jansen:2017oag,Jaramillo:2020tuu}. Among these methods, the continuous fraction method and the pseudo-spectral method are easily generalized to the case of rotation. Two dimensional pseudo-spectral methods are applied to solve the QNMs spectra~\cite{Xiong:2024urw,Cai:2025irl,Assaad:2025nbv}, which will be used in this work. By combining the hyperboloidal framework~\cite{Ansorg:2016ztf,PanossoMacedo:2018hab,PanossoMacedo:2019npm,Jaramillo:2020tuu,PanossoMacedo:2023qzp,PanossoMacedo:2024nkw,Zhou:2025xta,Cai:2025irl,Assaad:2025nbv,Ripley:2022ypi}, we cast the QNMs problem into a two-dimensional eigenvalue problem for each azimuthal mode, given $m$. The hyperboloidal framework removes the necessity of imposing external boundary conditions, since the time coordinate is naturally adapted to the causal structure of the black hole and the radiation zone. When solving the QNMs problem, the QNMs boundary conditions are built into the ``bulk'' of the operator, and no additional boundary conditions are required. The outward orientation of light cones at the computational domain's periphery simplifies boundary prescriptions to mandating solely a regular solution a condition effortlessly met in numerical implementations. The general construction of hyperboloidal coordinates for black hole spacetimes is introduced in~\cite{Zenginoglu:2007jw}, and the work~\cite{Zenginoglu:2011jz} suggests studying QNMs in the frequency domain on hyperboloidal foliations, while QNMs have been calculated on hyperboloidal foliations in time domain~\cite{Zenginoglu:2008uc, Zenginoglu:2009ey}. The numerical simulation of the Kerr black hole with massless scalar wave equation on hyperboloidal foliations is achieved in~\cite{Zenginoglu:2009hd}. The hyperboloidal framework has been achieved in many black holes, such as Schwarzschild black hole~\cite{Jaramillo:2020tuu}, RN black hole~\cite{Destounis:2021lum}, quantum corrected Schwarzschild black hole~\cite{Cao:2024oud}, Hayward black hole~\cite{Wu:2024ldo}, Boulware-Deser-Wheeler black hole~\cite{Cao:2024sot} and hairy black hole~\cite{Yang:2025hqk}. Note that when the master equation has a correction term of the velocity, the corresponding hyperbolidal framework can still work~\cite{Cao:2025qws}. See also some comments on the hyperboloidal framework for black hole QNMs~\cite{Shen:2025nbq}.

\begin{figure}[htbp]
	\centering
	\includegraphics[width=0.5\textwidth]{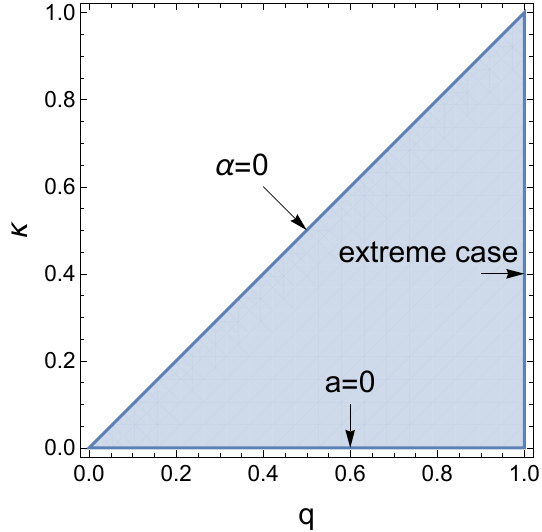}
	\caption{The parameter space $(q,\kappa)$ for the RQCBH, where the feasible range of parameters is represented by shaded areas.}
	\label{parameter_region}
\end{figure}

Before getting the QNMs, one should do some parameterizations for the rotating quantum corrected black hole in order to obtain the convenience of applying hyperboloidal coordinate transformation. Similar treatments can be found in some studies~\cite{Cao:2024oud,Wu:2024ldo,PanossoMacedo:2019npm,Destounis:2021lum,Zhou:2025xta}. According to the form of $\Delta$ in Eqs. (\ref{Delta_and_Sigma}), the function $\Delta$ can be factorized into
\begin{eqnarray}\label{Delta_factorized}
	\Delta(r)=\Big(1-\frac{r_{+}}{r}\Big)\Big(1-\frac{r_{-}}{r}\Big)(r^2+b_0r+c_0)\, .
\end{eqnarray}
 For convenience, $r_{+}$ denotes the event horizon and $r_{-}$ denotes the inner horizon. Comparing Eq. (\ref{Delta_factorized}) with Eqs. (\ref{Delta_and_Sigma}) and solving $\{b_0,c_0,M,\alpha\}$ from $\{r_{+},r_{-},a\}$, one obtains
\begin{eqnarray}\label{b0_c0_M_alpha}
	&&b_0=\frac{(r_{+}+r_{-})(r_{+}r_{-}-a^2)}{r_{+}^2+r_{+}r_{-}+r_{-}^2}\, ,\quad c_0=\frac{r_{+}r_{-}(r_{+}r_{-}-a^2)}{r_{+}^2+r_{+}r_{-}+r_{-}^2}\, ,\nonumber\\
	&&M=\frac{(r_{+}+r_{-})(a^2+r_{+}^2+r_{-}^2)}{2(r_{+}^2+r_{+}r_{-}+r_{-}^2)}\, ,\quad \alpha=\frac{4r_{+}^2r_{-}^2(r_{+}r_{-}-a^2)(r_{+}^2+r_{+}r_{-}+r_{-}^2)}{(r_{+}+r_{-})^2(a^2+r_{+}^2+r_{-}^2)^2}\, . 
\end{eqnarray}
Define two dimensionless parameters as follows
\begin{eqnarray}\label{kappa_q_definitions}
	\kappa:=\frac{a}{r_{+}}\, ,\quad q^2:=\frac{r_{-}}{r_{+}}\, ,
\end{eqnarray}
four quantities in Eqs. (\ref{b0_c0_M_alpha}) can be expressed by $\{\kappa,q,r_{+}\}$, i.e.,
\begin{eqnarray}\label{b0_c0_M_alpha_in_terms_of_kappa_q_rp}
	&&b_0=\frac{(1+q^2)r_{+}(q^2-\kappa^2)}{1+q^2+q^4}\, ,\quad c_0=\frac{q^2r_{+}^2(q^2-\kappa^2)}{1+q^2+q^4}\, ,\nonumber\\
	&&M=\frac{(1+q^2)r_{+}(1+q^4+\kappa^2)}{2(1+q^2+q^4)}\, ,\quad \alpha=\frac{4q^4(1+q^2+q^4)r_{+}^2(q^2-\kappa^2)}{(1+q^2)^2(1+q^4+\kappa^2)^2}\, .
\end{eqnarray}
Therefore, the function $\Delta$ becomes
\begin{eqnarray}
	\frac{\Delta}{r_{+}^2}=\Big(1-\frac{r_{+}}{r}\Big)\Big(1-\frac{r_{-}}{r}\Big)\Big[\Big(\frac{r}{r_{+}}\Big)^2+\frac{(1+q^2)(q^2-\kappa^2)}{1+q^2+q^4}\Big(\frac{r}{r_{+}}\Big)+\frac{q^2(q^2-\kappa^2)}{1+q^2+q^4}\Big]\, .
\end{eqnarray}
From definitions (\ref{kappa_q_definitions}), the feasible parameter space $(q,\kappa)$ is displayed in Fig. \ref{parameter_region}, where the shaded part is an isosceles right angled triangle. In Fig. \ref{parameter_region}, the line $\kappa=q$ corresponds to the Kerr black hole, the line $\kappa=0$ corresponds to the spherical quantum corrected black hole~\cite{Lewandowski:2022zce}, and the line $q=1$ corresponds to the extreme black hole. Note that for $\kappa=q$, such parameterization approach is consistent with~\cite{PanossoMacedo:2019npm}. 

Now, it is ready to solve the QNMs spectra of RQCBH within the hyperboloidal framework. The complete mapping from Boyer-Lindquist to the hyperboloidal coordinates are given by
\begin{eqnarray}\label{hyperboloidal_coordinates}
	t=\lambda\Big[\tau-h(\sigma,\theta)\Big]-r_{\star}(r(\sigma))\, ,\quad r(\sigma)=\lambda\frac{\rho(\sigma)}{\sigma}\, ,\quad \varphi=\phi-k(r(\sigma))\, ,
\end{eqnarray}
where $\lambda$ is the characteristic scale, $h(\sigma,\theta)$ is called the height function, and $k(r(\sigma))$ is the phase function. The hyperboloidal coordinates of Kerr black hole have been work out in~\cite{PanossoMacedo:2019npm}. Here, we use the approach from~\cite{PanossoMacedo:2019npm} to solve the hyperboloidal coordinates for RQCBH, with the most important being to solve the height function. In Appendix \ref{height_functions}, the height function is derived. If we apply the minimal gauge (radial function fixing gauge) and choose the length scale as $\lambda=r_{+}$, and get
\begin{eqnarray}
	\rho(\sigma)=1\, ,\quad \beta(\sigma)=1\, ,\quad h(\sigma,\theta)=-\frac{2}{\sigma}+\frac{2(1+q^2)(1+q^4+\kappa^2)}{1+q^2+q^4}\ln\sigma\, ,
\end{eqnarray}
where we have used Eqs. (\ref{b0_c0_M_alpha_in_terms_of_kappa_q_rp}). As a result, the function $\Delta$ can be expressed as
\begin{eqnarray}
	\frac{\Delta(\sigma)}{r_{+}^2}=(1-\sigma)(1-q^2\sigma)\Big[\frac{1}{\sigma^2}+\frac{(1+q^2)(q^2-\kappa^2)}{(1+q^2+q^4)\sigma}+\frac{q^2(q^2-\kappa^2)}{1+q^2+q^4}\Big]\, .
\end{eqnarray}

Finally, the master function $\Phi(\tau,\sigma,\theta,\phi)$ is written as~\cite{PanossoMacedo:2019npm,Assaad:2025nbv}
\begin{eqnarray}\label{U_decomposition}
	\Phi(\tau,\sigma,\theta,\phi)=\Omega\sum_{m=-\infty}^{\infty}\cos^{\delta}(\theta/2)\sin^{\delta}(\theta/2)V_m(\tau,\sigma,\theta)\mathrm{e}^{\mathrm{i}m\phi}\, ,
\end{eqnarray}
with the exponent $\delta=|m|$. The angle $\phi$-dependence is $\mathrm{e}^{\mathrm{i}m\phi}$, in which $m$ is the azimuthal number. With the subsitution $x=\cos\theta$, one can achieve the final form. At the same time, we perform a first-order reduction in time, i.e., $W_m=\partial_\tau V_m$, and Eq. (\ref{KG_BL}) in the minimal gauge can be rewritten as two partial differential equations, involving first-order derivative respect to time and second-order derivative respect to $\sigma$ and $x$, i.e.,
\begin{eqnarray}\label{dynamics_eq}
	\partial_\tau \begin{bmatrix}
		V_m(\tau,\sigma,x)\\
		W_m(\tau,\sigma,x)
	\end{bmatrix}=\mathrm{i}L\begin{bmatrix}
		V_m(\tau,\sigma,x)\\
		W_m(\tau,\sigma,x)
	\end{bmatrix}\, ,
\end{eqnarray}
where the operator $L$ is defined as
\begin{eqnarray}\label{operator_L}
	L=\frac{1}{\mathrm{i}}
	\begin{bmatrix}
		0 & 1\\
		L_1 & L_2
	\end{bmatrix}\, .
\end{eqnarray}
We call $L$ is the time generator of the linear dynamics for the RQCBH, and we explicitly write out the parameters on which the operator $L$ depends. The expressions of the operators $L_1$ and $L_2$ can be found in the Appendix \ref{expressions_L1_and_L2}. Consider the time part $ \mathrm{e}^{\mathrm{i}\omega\tau}$, we arrive at the two-dimensional eigenvalue problem as follow
\begin{eqnarray}\label{QNM_eigenvalue_problem}
	L\begin{bmatrix}
		\mathbb{V}_m(\omega,\sigma,x)\\
		\mathbb{W}_m(\omega,\sigma,x)
	\end{bmatrix}=\omega
	\begin{bmatrix}
		\mathbb{V}_m(\omega,\sigma,x)\\
		\mathbb{W}_m(\omega,\sigma,x)
	\end{bmatrix}\, ,
\end{eqnarray}
in which boundary conditions are encoded in $L$. Within the help of hyperboloidal framework, the eigenfunctions of the spectral problem are regular in the region $(\sigma,x)\in[0,1]\times[-1,1]$. Note that the divergence of the angular part has been excluded by the term $\sin^{|m|}(\theta/2)\cos^{|m|}(\theta/2)$. We will use a two-dimensional pseudo-spectral method to solve the eigenvalue problem in a tensor product grid. The details of such a method can be found in~\cite{Jansen:2017oag,Jaramillo:2020tuu,Xiong:2024urw,Cai:2025irl,doi:10.1137/1.9780898719598,Miguel:2023rzp,Assaad:2025nbv}. For each direction grid, the Chebyshev-Lobatto grid is used, where the Chebyshev-Lobatto grid is given by
\begin{eqnarray}
    \sigma_i&=&\frac{1}{2}\Big[1+\cos\Big(\frac{i\pi}{N_\sigma}\Big)\Big]\, ,\quad i=0,1,\cdots,N_\sigma-1,N_\sigma\ ,\nonumber\\
    x_j&=&\cos\Big(\frac{j\pi}{N_x}\Big)\, ,\quad j=0,1,\cdots,N_x-1, N_x\, ,
\end{eqnarray}
and $N_\sigma$ is the resolution for the $\sigma$ direction, $N_x$ is the resolution for the $x$ direction. Therefore, the dimension of the finite rank approximation $\mathbf{L}$, a square matrix, for the original operator $L$ is $2(N_\sigma+1)(N_x+1)\times2(N_\sigma+1)(N_x+1)$. One can refer to Appendix \ref{numerical_accuracy_test} about the convergence test with different resolutions. In our present work, the resolutions $N_\sigma=20$ and $N_x=10$ are chosen. An important distinction between the spectra obtained from the two-dimensional eigenvalue problem and those from the one-dimensional problem (spherically symmetric) lies in the fact that the two-dimensional results encompass spectra of QNMs for all the angular momentum numbers $\ell\ge|m|$. Note that for $\kappa=0$, the metric reduces to the spherically symmetric case and the QNMs spectra have been extensively studied in~\cite{Gong:2023ghh,Yang:2022btw,Cao:2024oud} and their results can be used to validate our QNMs spectra results within $a=0$.

Because we will use GW data for parameter estimation in the followings, we are concerned about the case of $m=2$. Given a rotating quantum corrected black hole with the mass $M$, the spin $a$ and the quantum correction parameter $\alpha$, there exists an infinite number of QNMs spectra labeled $\omega^{\pm}_{\ell mn}=2\pi f^{\pm}_{\ell mn} +\mathrm{i}/\tau^{\pm}_{\ell mn}$, where $\ell$ and $m$ are the usual angular indices, and $n$ is called the overtone number from $0$. The superscript $\pm$ indicates the sign of the real part of the spectrum. The modes in right half-plane are called the ``regular'' modes, while the modes in left half-plane are called the ``mirror'' modes. There is a symmetry between the spectra of the ``regular'' and ``mirror'' modes given by $\omega^{-}_{lmn}=-\omega^{+\star}_{l-mn}$, in which $\star$ represents the complex conjugate.

\begin{figure}[htbp]
	\centering
	\includegraphics[width=0.45\textwidth]{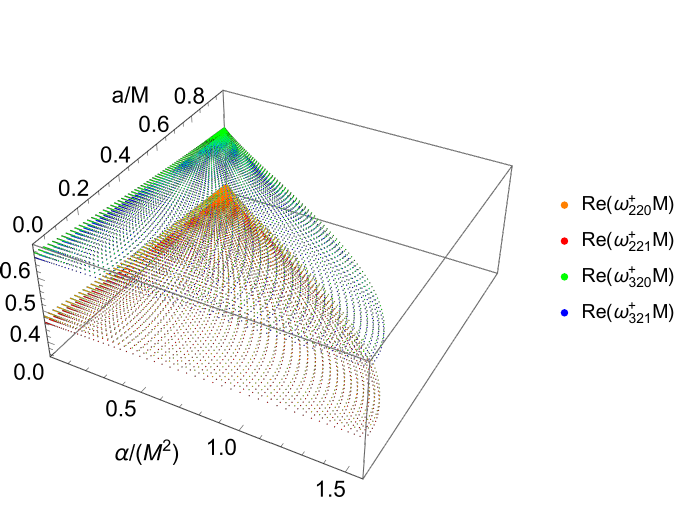}\hspace{1.5cm}
    \includegraphics[width=0.45\textwidth]{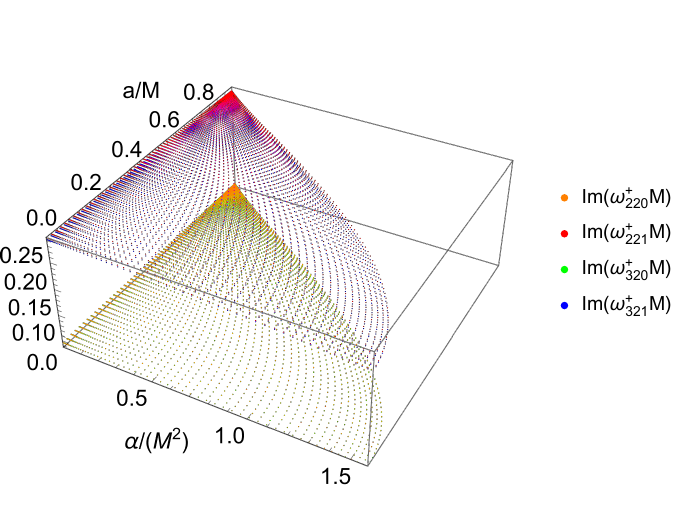}
    \includegraphics[width=0.45\textwidth]{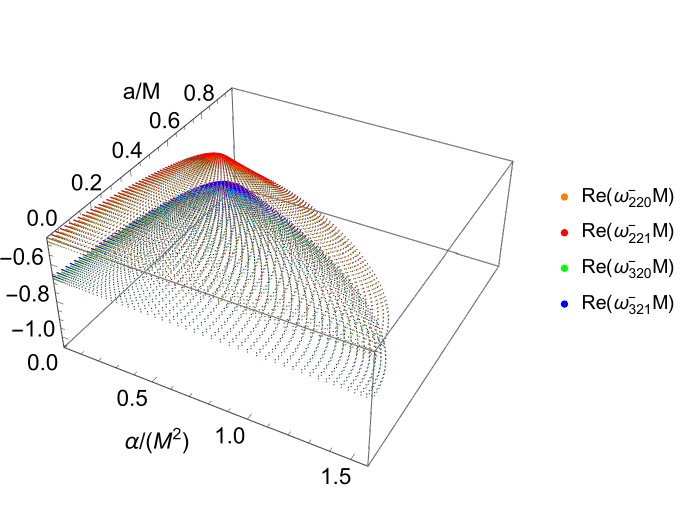}\hspace{1.5cm}
    \includegraphics[width=0.45\textwidth]{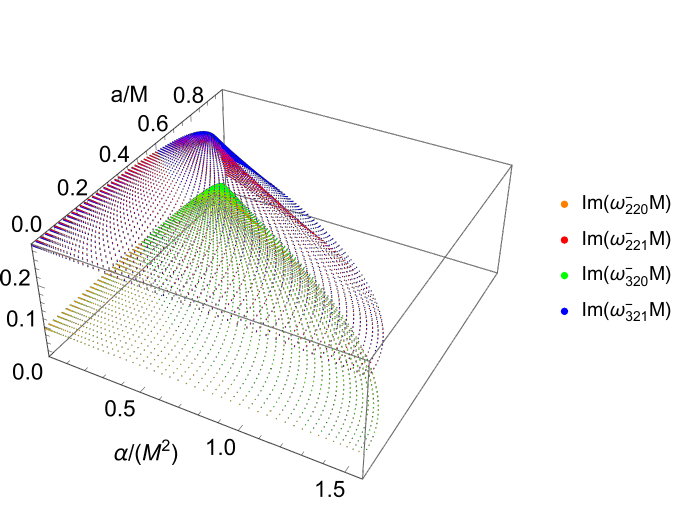}
	\caption{The QNMs spectra for RQCBH with $\ell=3$, $m=2$ and $\ell=2$, $m=2$. The left top panel is the real part of the four ``regular'' modes. The right top panel is the imaginary part of the four ``regular'' modes. The left bottom panel is the real part of the four ``mirror'' modes. The right bottom is the imaginary part of the four ``mirror'' modes.}
	\label{Spectra_t}
\end{figure}

So far, we have obtained the QNMs spectra within the hyperboloidal framework. However, considering that in the processing of GW data, the spectra in B-L coordinate are used, so we need to convert the results of the spectra in the hyperboloidal coordinate into the results in the B-L coordinate. For simplicity, the points $(q,\kappa)$ in the $q-\kappa$ plane are homogeneous, in which the points are confined to the set $\{(q,\kappa)|0\le\kappa\le q\, ,q\le0.9\}$, and $\Delta \kappa=\Delta q=0.01$. Continuously, from Eqs. (\ref{b0_c0_M_alpha_in_terms_of_kappa_q_rp}), we have
\begin{eqnarray}\label{alpha_M_and_a_M}
    \frac{\alpha}{M^2}=\frac{16 q^4(q^4+q^2+1)^3(q^2-\kappa ^2)}{(q^2+1)^4(\kappa ^2+q^4+1)^4}\, ,\quad \frac{a}{M}=\frac{2\kappa(q^4+q^2+1)}{(q^2+1)(\kappa ^2+q^4+1)}\, .
\end{eqnarray}
Using Eqs. (\ref{alpha_M_and_a_M}), we finally get the samples in the $\bar{\alpha}-\bar{a}$ plane, which are shown in Fig. \ref{Samples_alpha_a}. The mass $M$ of the black hole acts simply as an overall scale on the QNMs spectra, so for simplicity, we assume that the unit is $M=1$. According to the relation ${}^{[t]}\omega={}^{[\tau]}\omega/r_{+}$ [see Eqs. (\ref{hyperboloidal_coordinates})], we display the spectra on each points $(\bar{\alpha},\bar{a})$ for the modes with $\{\ell=2, m=2, n=0\}$, $\{\ell=2, m=2, n=1\}$, $\{\ell=3, m=2, n=0\}$, $\{\ell=3, m=2, n=1\}$ of the RQCBH including the ``regular'' modes and the ``mirror'' modes in Fig. \ref{Spectra_t}.

\section{Bayesian analysis and parameter estimation}\label{Bayesian_analysis}
In this section, the Bayesian analysis and parameter estimation will be done. We begin to establish the functions $\omega_{\ell mn}(M,\bar{a},\bar{\alpha})$ in terms of the parameters $M$, $a$ and $\alpha$ by using the fitting method. Note that for this time, such $\omega$ should be obtained from the B-L coordinate. In the case of non spin and no quantum correction, these functions reduce to the Schwarzschild QNMs spectra. In~\cite{Nagar:2018zoe}, the authors fit the spectra by using a rational function, and the same fitting method is applied to Kerr-Newman (K-N) black hole~\cite{Carullo:2021oxn}. The only difference is that for the K-N case, the fitting function becomes a bivariate rational fraction function. Motivated by the above and consider that there are also two independent free parameters in the present model, we use the bivariate rational fraction function, namely
\begin{eqnarray}\label{fitting_model}
    \mathbf{X}=\mathbf{X}_0\frac{1+\sum_{1\le k+j\le N_1}b_{k,j}\bar{\alpha}^{k}\bar{a}^{j}}{1+\sum_{1\le k+j\le N_2}c_{k,j}\bar{\alpha}^{k}\bar{a}^{j}}\, ,\quad k\in\mathbb{N}\, ,j\in\mathbb{N}\, ,N_1\in\mathbb{N}^{+}\, ,N_2\in\mathbb{N}^{+}\, ,
\end{eqnarray}
to fit obtained spectra data. Here, some notation descriptions should be given. The symbol $\mathbf{X}$ corresponds to $\omega$, while the symbol $\mathbf{X}_0$ stands for the Schwarzschild value of the corresponding quantity. In other words, as $\bar{\alpha}=\bar{a}=0$, we have $\mathbf{X}=\mathbf{X}_0$. The notations $\bar{\alpha}$ and $\bar{a}$ are defined as $\bar{\alpha}=\alpha/M^2$ and $\bar{a}=a/M$, which has been mentioned in Sec. \ref{RQCBH}. Furthermore, we think that in Eq. (\ref{fitting_model}), the order of the bivariate polynomial in the numerator can different from that in the denominator. Therefore, their orders are recorded as $N_1$ and $N_2$ respectively. But in practice, we take $N_1=N_2=4$. We do these fittings on \textit{Mathematica} by using the built-in function \textit{NonlinearModelFit}, where the starting values of all parameters $\{b_{k,j}, c_{k,j}\}$ to be fitted are set to $0$. It can be found that the maximum residuals of the QNM spectra fittings in Appendix \ref{numerical_accuracy_test}.

\begin{figure}[htbp]
    \centering
    \includegraphics[width=0.43\linewidth]{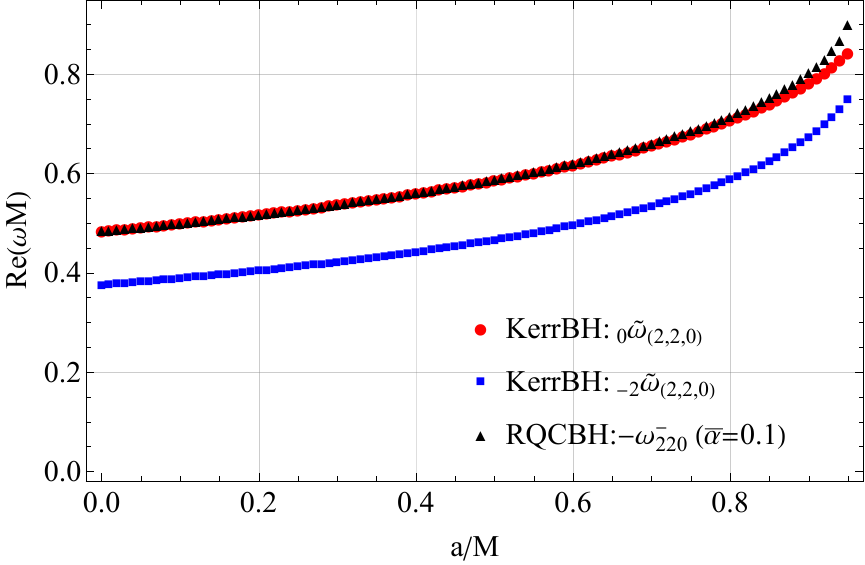}
    \hfill
    \includegraphics[width=0.45\linewidth]{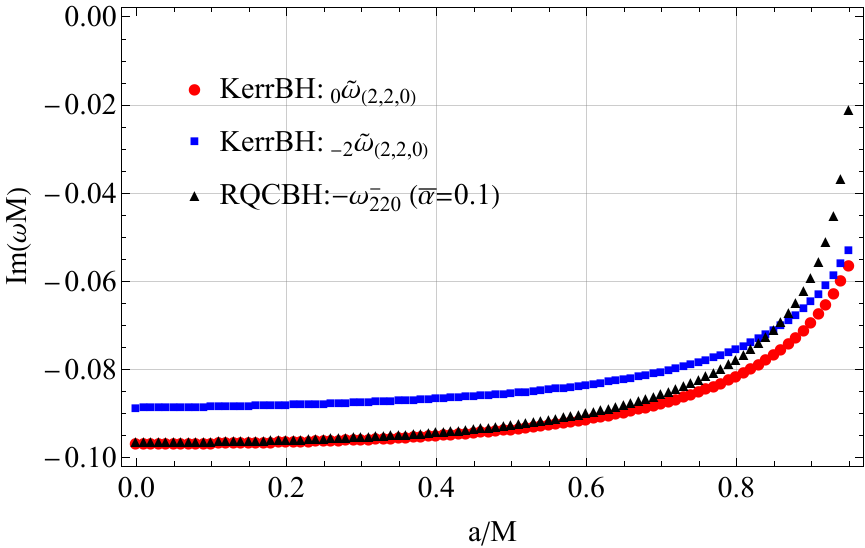}
    \caption{The real (left) and imaginary (right) parts of QNMs spectra as functions of the dimensionless spin parameter $a/M$. The red circles and blue squares correspond to the Kerr QNMs spectra $_{0}\tilde{\omega}_{(2,2,0)}$ and $_{-2}\tilde{\omega}_{(2,2,0)}$, respectively. The black triangles represent the RQCBH mode $-\omega^{-}_{220}$ with the fixed parameter $\bar{\alpha}=0.1$. }
    \label{fig: comparison between Kerr and RQC}
\end{figure}

Now, we employ \texttt{pyRing} for the subsequent analysis. The Bayesian analysis based on the time-domain data is subject to obtaining the posterior distribution $p(\bm{\theta} | d)$, where $\bm{\theta}$ represent the set of waveform model parameters, and $d$ represents the observation data. According to Bayes' theorem, $p(\bm{\theta} | d)$ is determined by $p(\bm{\theta} | d)=\mathcal{L}(d | \bm{\theta}) \pi(\bm{\theta})/\mathcal{Z}$, where the $\mathcal{L}(d | \bm{\theta})$ is the likelihood function, $\pi(\bm{\theta})$ is the prior distribution, and $\mathcal{Z} = \int \mathrm{d} \bm{\theta} \mathcal{L}(d | \bm{\theta}) \pi(\bm{\theta}) $ refers to the ``evidence''. For ground-based gravitational-wave detectors, the ringdown signal decays very rapidly, and the noise of the whitened data in this regime is therefore well modeled as Gaussian with zero mean. The likelihood function is constructed on this assumption. The whitened data needs the waveform model and the Power Spectral Density (PSD). The PSD is computed with the standard Welch method. More importantly, the waveform model determines which parameters $\bm{\theta}$ of gravitational theory are inferred.

\begin{table}[htbp]
    \centering
   \renewcommand{\arraystretch}{2.0} 
    \begin{tabularx}{\textwidth}{c | c c c c c c c c c }
        \hline
        \hline
        GW events & fit modes $(\ell,m,n)$ & peak time  $(t_{H_{1}})$ & start time $(t_{0})$ & $M$ & $\bar{a}$ & $\mathrm{ra~(rad)}$ & $\mathrm{dec~(rad)}$ & $\mathrm{\psi~(rad)}$ & $\cos \iota$  \\
        \hline
        $\mathrm{(i)}$ GW150914 & $(2,2,0)+(2,2,1)$ & $126259462.4232$ & $+0 \mathrm{ms}$ & $[30,100]$ & $[0.1,0.9]$ & 1.95 & -1.27 & 0.82 & -1 \\
        $\mathrm{(ii)}$ GW190521 & \makecell{$(2,2,0)+(2,2,1)$ \\$+(3,2,0)$} & $1242442967.4306$ & \makecell{$+12.7~\mathrm{ms}$} & $[200,400]$ & $[0.1,0.9]$ & $0.16$ & $-1.14$ & $2.38$ & $0.67$ \\
        $\mathrm{(iii)}$ GW231123 & $(2,2,0)+(2,2,1)$ & $1384782888.5998$ & $+ 32.3 \mathrm{ms}$ & $[200, 400]$ & $[0.1,0.9]$ & $4.03$ & $0.73$ & $2.54$ & $0.87$ \\
        \hline
        \hline
    \end{tabularx}
    \caption{Preset for Bayesian inference using \texttt{pyRing}.}
    \label{tab: the preset for pyRing}
\end{table}

In \texttt{pyRing}, the standard Kerr waveform model is constructed by~\cite{Carullo:2019flw,Lim:2019xrb}
\begin{eqnarray}
    h_{+}+\mathrm{i} h_{\times}=\frac{M_f}{D_L} \sum_{\ell=2}^{\infty} \sum_{m=-\ell}^{+\ell} \sum_{n=0}^{\infty}(h_{\ell m n}^{+}+h_{\ell m n}^{-})\, ,
\end{eqnarray}
with
\begin{eqnarray}
    h_{\ell m n}^{+} &=&A_{\ell m n}^{+} S_{\ell m n}(\iota, \varphi) \mathrm{e}^{\mathrm{i}\left[\left(t-t_{\ell m n}\right) \tilde{\omega}_{\ell m n}+\phi_{\ell m n}^{+}\right]} \, , \\
    h_{\ell m n}^{-} &=&A_{\ell m n}^{-} S_{\ell-m n}(\iota, \varphi) \mathrm{e}^{-\mathrm{i}\left[\left(t-t_{\ell m n}\right) \tilde{\omega}_{\ell m n}^{\star}-\phi_{\ell m n}^{-}\right]}\, .
\end{eqnarray}
Here, $\tilde{\omega}_{\ell m n} = 2\pi f_{\ell m n} + \mathrm{i}/\tau_{\ell m n}$ is the Kerr QNMs spectra, which can be expressed as a function $\tilde{\omega}_{\ell m n}(M_{f}, a_{f})$, and $\star$ represents complex conjugation. The amplitudes $A^{+/-}_{\ell m n}$ and phases $\phi^{+/-}_{\ell m n}$ characterize the excitation of each modes. The inclination of the BH final spin relative to the observer’s line of sight is denoted by $\iota$, while $\varphi$ corresponds to the azimuthal angle of the line of sight in the BH frame. $S_{\ell m n}$ are the spin-weighted spheroidal harmonics and $t_{\ell m n}=t_{0}$ is a reference start time~\cite{pyRingDocs2025}. 

It should be noted that the waveform templates implemented in \texttt{pyRing} are constructed based on perturbations with spin weight $s = -2$, whereas in this work we employ the $s = 0$ QNMs to infer parameters' posterior distribution. In addition, taking the $s = 0$ data from~\cite{Berti2025RingdownData,Berti:2009kk} as the reference, the Kerr QNMs spectra $_{0}\tilde{\omega}_{|\text{KerrBH}}$ with $(\ell, m, n) = (2, 2, 0)$ corresponds to the QNMs spectra in the RQCBH model $-_{0}\omega^{-}_{|\text{RQCBH}}$ with $(2, 2, 0)$. Based on this correspondence, we compare the Kerr QNMs spectra for the $(2,2,0)$ modes with $s=0$ and $s=-2$ perturbations against the RQCBH $(2,2,0)$ mode with $s=0$ perturbation to highlight their differences, as shown in Fig. \ref{fig: comparison between Kerr and RQC}. In the implementation of \texttt{pyRing}, the substitution of the QNMs in the waveform template,
\begin{equation}
    _{s=-2}\tilde{\omega}_{(2,2,0)|\text{KerrBH}}  \rightarrow -_{s=0}\omega_{(2,2,0)|\text{RQCBH}}^{-} \, , 
\end{equation}
is carried out within the inference pipeline for the RQCBH model. 

We use three GW events, (i)~GW150914~\cite{LIGOScientific:2016aoc, LIGOScientific:2016vlm, Cotesta:2022pci}, (ii)~GW190521~\cite{LIGOScientific:2020ufj, Siegel:2023lxl}, and (iii)~GW231123~\cite{LIGOScientific:2025rsn}, to construct the Bayesian inference pipline, with their corresponding \texttt{pyRing} configurations summarized in Tab. \ref{tab: the preset for pyRing}. For the events (i) and (iii), we employ the fundamental mode together with the first overtone to infer the parameters' posterior function. For the event (ii), given its potential to reveal higher-order contributions, we additionally include the $(\ell,m,n)=(3,2,0)$ mode. Considering the improvement in detector sensitivity and the higher signal-to-noise ratios of the two events, the ringdown start times for all cases except (i) are shifted further to avoid contamination from nonlinear dynamics. The prior of the quantum correction parameter $\bar{\alpha}$ is set to $(0, 27/16)$ uniformly, and its values that do not satisfy the condition with Eq. (\ref{eq: constrain on alpha and a}) are excluded during the Markov-Chain-Monte-Carlo (MCMC) sampling process. To accelerate the computation, we fix the parameters of sky location, polarization angle $\psi$, and inclination parameter $\cos \iota$ to the maximum-likelihood value corresponding to their inspiral-merger-ringdown (IMR) information. For case (i), the IMR data are obtained from \textsc{IMRPhenomP}~\cite{Hannam:2013oca}, while for cases (ii) and (iii), they are taken from \textsc{NRSur7dq4}~\cite{Varma:2019csw}. The sampler settings used are: $4096$ live points, $4096$ as a maximum number of MCMC internal steps, and a pool-size composed of $100$ walkers. The stopping condition is set by requiring an estimated precision of $0.1$ on the logarithm of the evidence.

We employed two inference models in our analysis. The first model adopts uniform priors, with the corresponding mass and spin distributions listed in Tab. \ref{tab: the preset for pyRing}. The second model introduces informative priors predicted from the inspiral–merger phase. This approach effectively imposes Gaussian prior constraints on the mass and spin parameters, where the distributions inferred from the inspiral–merger stage provide a physically well-motivated and reasonable realization of such Gaussian priors. Technically, we read the posterior distributions of mass and spin from the corresponding IMR analysis as the prior distributions on the ringdown analysis. Since the complete IMR waveform is generated from the physical parameters of the binary system (such as the mass ratio and the component spins), the ringdown portion of the IMR waveform does not explicitly take the mass and spin of the remnant black hole as independent parameters. Therefore, this procedure does not involve any double use of the observational data. In addition, We also incorporate the standard Kerr-based inference results from IMR for comparison.

\begin{figure}[htbp]
    \centering
    \includegraphics[width=0.7\linewidth]{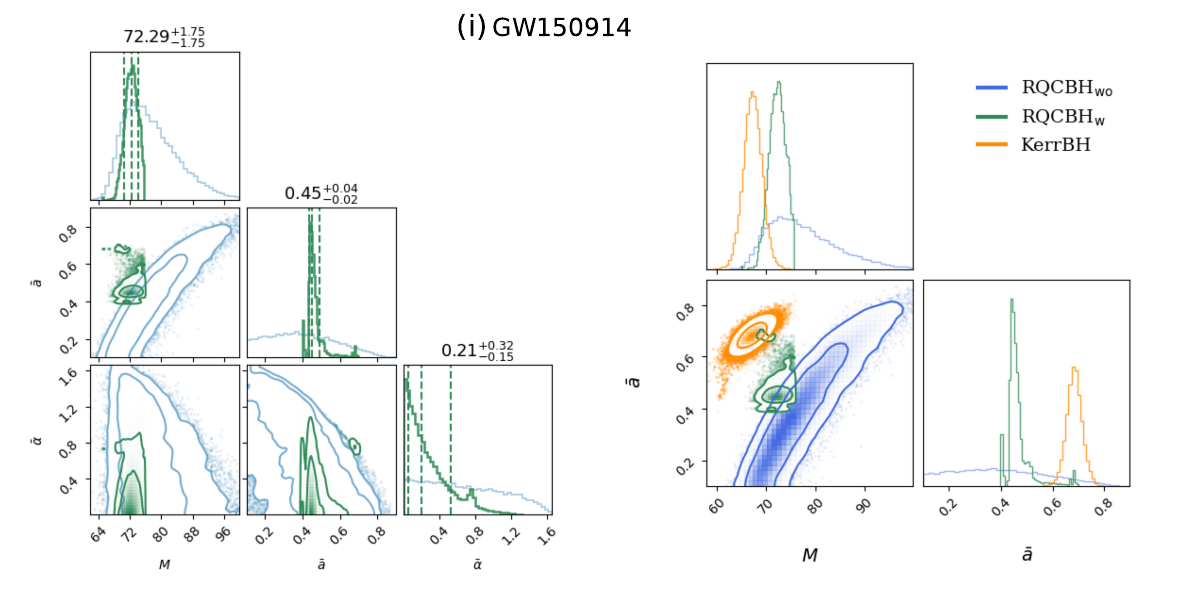}
    \includegraphics[width=0.7\linewidth]{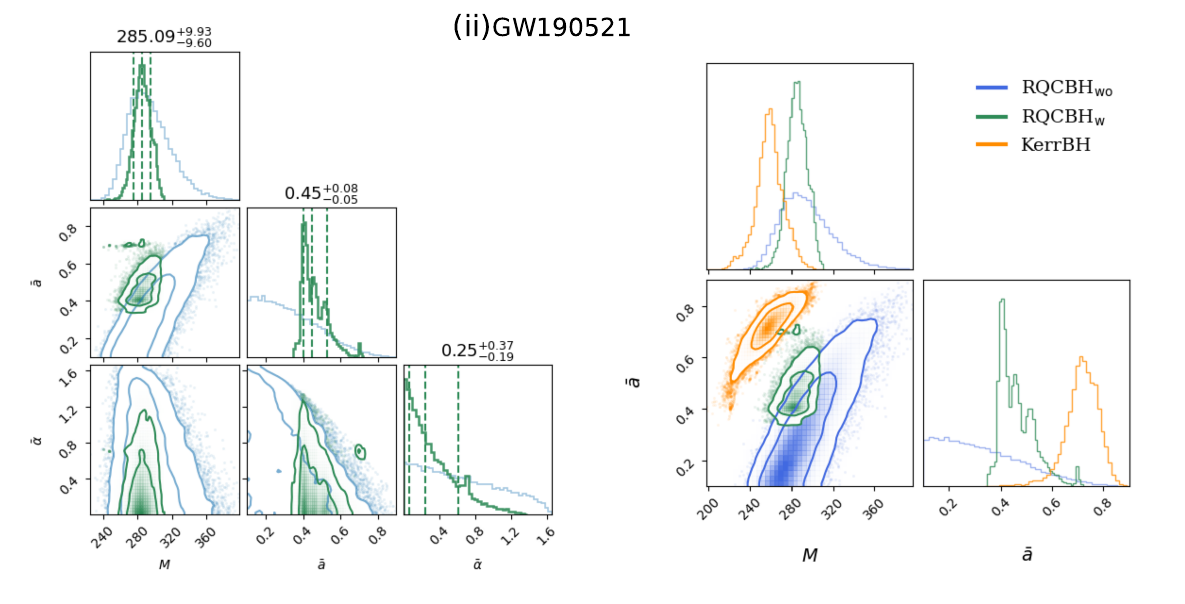}
    \includegraphics[width=0.7\linewidth]{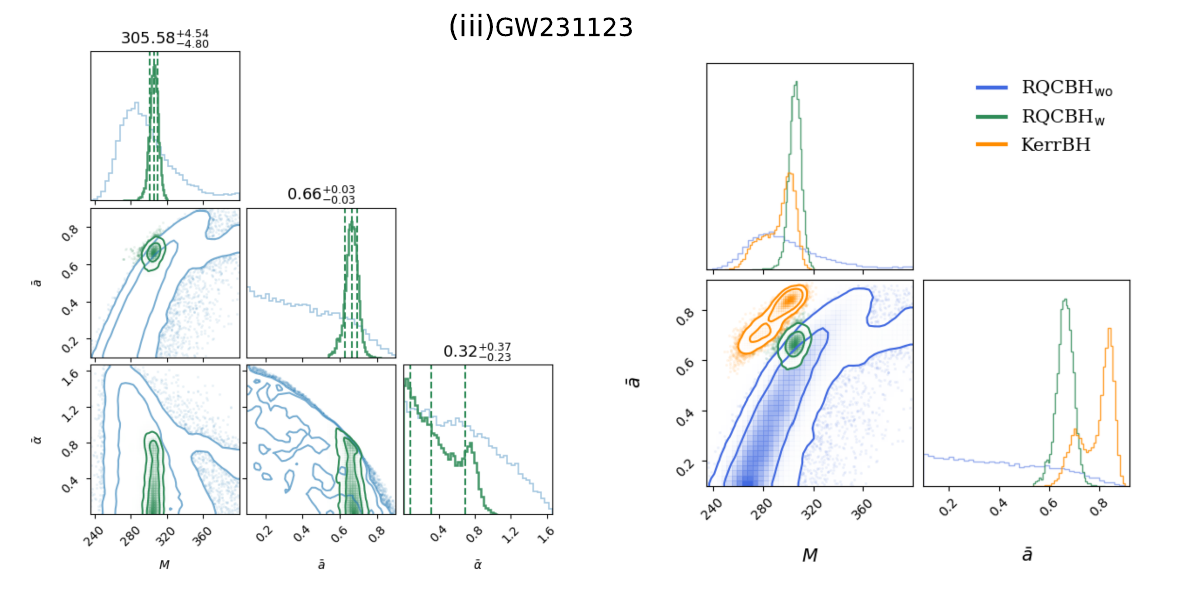}
    \caption{Posterior distributions of the RQCBH parameters inferred from ringdown signals for $\mathrm{(i)}$ GW150914 and $\mathrm{(ii)}$ GW190521, and $\mathrm{(iii)}$ GW231123. Left three panels: corner plots show one-dimensional and two-dimensional posteriors of the mass $M$, spin $\bar{a}$, and the quantum correction parameter $\bar{\alpha}$ for the RQCBH model. The blue curves correspond to the RQCBH model without informative priors, and the green curves represent the RQCBH model with informative priors. Contours correspond to $68\%$ and $95\%$ credible regions, and vertical dashed lines indicate $16$th, $50$th, and $84$th percentiles. Right three panels: the comparison of posterior distributions of the mass $M$ and spin $\bar{a}$, including those obtained from the IMR analysis based on the Kerr assumption, which is denoted by yellow curves. The contours in the two-dimensional plots also denote $68\%$ and $95\%$ credible regions.}
    \label{fig:inference results}
\end{figure}

The results are shown in Fig. \ref{fig:inference results}. This figure presents a systematic comparison of the posterior distributions obtained from parameter inference on the ringdown signals under different model assumptions. Each horizontal row displays the inference results acquired by using the respective event data, labeled in each upper corner. The left three panels display the inference results with the quantume corrected parameter $\bar{\alpha}$ using \texttt{pyRing} for different two different inference models, the blue curves correspond to the RQCBH model without informative priors, and the green curves represent the RQCBH model with informative priors. Contours correspond to $68\%$ and $95\%$ credible regions, and vertical dashed lines indicate $16$th, $50$th, and $84$th percentiles. While the three right panels compare the posterior results of the $M$ and $\bar{a}$, including those obtained from the IMR analysis based on the Kerr assumption, which is denoted by yellow curves. From the overall inference results, the RQCBH model exhibits a wide posterior spread in the absence of additional informative priors. The correlation between the quantum corrected parameter $\bar{\alpha}$ and the spin $\bar{a}$ primarily reflects the restriction relationship, as indicated by Fig. \ref{Samples_alpha_a}, offering only weak limits on $\bar{\alpha}$. By incorporating informative priors predicted from the inspiral–merger phase, the constraint on $\bar{\alpha}$ is substantially improved. Comparing the three different models, the estimated black hole masses remain largely similar. Notably, once informative priors are included, the spin inferred from the RQCBH model begins to be significant and differ from that of the Kerr model. This change is a direct consequence of the informative priors and the quantum-correction parameter $\bar{\alpha}$ that alters the spacetime structure and produces a detectable shift in the inferred spin.

It's notable that the inference results using the scalar QNMs spectra in \texttt{pyRing} pipeline should not be interpreted as accurate constraints on the RQCBH model. Furthermore, it would be important to investigate, once gravitational QNMs are obtained, how they differ from scalar QNMs, and to what extent parameter inference based on gravitational QNMs may differ from that based on scalar QNMs. We expect that the QNMs associated with gravitational and scalar perturbations should share similar qualitative properties, with their primary differences arising from the quantitative dependence on the black hole parameters. This expectation is analogous to the well-established relationship between scalar and gravitational QNMs in the Kerr spacetime. For the Kerr black hole, analysis indicates that although the two kinds of modes differ in their mathematical formulation, as indicated by Fig. \ref{fig: comparison between Kerr and RQC}, the posterior distributions of the black hole mass and spin inferred from different modes are remarkably similar, see Fig. 6 of Ref. \cite{Taylor:2024duw}. Therefore, as an initial investigation, adopting scalar QNMs in combination with GW data to perform parameter inference is still valuble. Finally, we would like to emphasize that the strong correlation between $\bar{\alpha}$ and $\bar{a}$ is not tied to the specific choice of perturbation. Instead, this correlation originates from intrinsic consistency requirements of the black hole spacetime itself. Thus, the same relation should persist for the case of the gravitational perturbation as well.

.

\section{Conclusions and discussion}\label{Conclusions_and_discussion}
In this work, we have systematically investigated the scalar QNMs spectra of rotating quantum corrected black hole (RQCBH), obtained by applying the Newman–Janis generating method to spherically symmetric black hole in the loop quantum gravity~\cite{Lewandowski:2022zce}. To compute the QNMs spectra with $s=0$ perturbation, we employ the hyperboloidal framework combined with a two-dimensional pseudo-spectral method, which naturally incorporates boundary conditions of QNMs at the event horizon and null infinity. The results of QNMs spectra are shown in Fig. \ref{Spectra_t}.  The resulting QNMs spectra are characterized by three physical parameters, namely the mass $M$, dimensionless spin $\bar{a}$, and dimensionless quantum correction parameter $\bar{\alpha}$, and it will reduce to the Kerr or quantum corrected Schwarzschild black hole in the $\bar{\alpha} \to 0$ or $\bar{a} \to 0$ limits. 

Based on these QNMs spectra, we constructed a Bayesian inference framework using the \texttt{pyRing} package to explore the influence of the quantum parameter $\bar{\alpha}$ on GW ringdown analyses.  First, the obtained spectra are fitted by bivariate rational functions in order to provide analytic expressions $\omega_{\ell m n}(M, \bar{a}, \bar{\alpha})$ for parameter estimation. By applying the RQCBH-based QNMs templates, both with and without informative priors of mass and spin, to representative events GW150914, GW190521, and GW231123, we find that the use of informative priors consistently yields a tighter posterior on $\bar{\alpha}$ compared to analyses without such priors. Besides, comparing the two cases discussed above together with the Kerr model shows that the inferred black hole mass parameters remain broadly consistent across all models. But once informative priors are included, the spin inferred from the RQCBH model begins to be significant and differs from that of the Kerr model, since the quantum-correction parameter $\bar{\alpha}$ alters the spacetime structure and produces a detectable shift in the inferred spin.

Notably, this pipeline employs scalar QNMs spectra, while the waveform model beyond Kerr black hole in $\texttt{pyRing}$ is designed for the tensor perturbation; therefore, the resulting posterior distributions should be interpreted as a methodological investigation rather than as physical constraints on $\bar{\alpha}$ from GW observations. The difficulty in computing gravitational QNMs fundamentally arises from the fact that the spherically symmetric quantum corrected metric proposed in Ref. \cite{Lewandowski:2022zce} lacks an underlying fundamental theory, action, and explicit field equations, which prevents a direct tensor-perturbation analysis. However, compared with the NJ algorithm, the GD method suggests that, at least at a formal level, there exists a corresponding Einstein-like field equation for the RQCBH, $ \tilde{G}_{\mu \nu}=\tilde{T}_{\mu\nu}$.  If such an explicit equation could be established, the gravitational QNMs can, in principle, be efficiently computed using the spectral and pseudo-spectral method~\cite{Lam:2025elw,Chung:2023wkd,Chung:2024ira} or modified Teukolsky formalism~\cite{Li:2022pcy,Hussain:2022ins, Cano:2023tmv, Cano:2023jbk}. This opens up the possibility of computing the gravitational QNMs by analyzing the tensor perturbation of the above field equation.

Several limitations of the present analysis should be noted. Although the numerical accuracies for the QNM spectra fittings and the convergence of the pseudo-spectral method have been given in Appendix \ref{numerical_accuracy_test}, these methods itself may lead to numerical errors for higher modes and overtones, as well as near the boundaries of the parameter domain. Nevertheless, these limitations do not affect the overall qualitative conclusions of this work. Besides, the replacement of scalar RQCBH QNMs in the Kerr $s=-2$ waveform model will introduce model mismatch~(including mode itself and spin-weighted spheroidal harmonics) and systematic uncertainties on Bayesian inference. A potential concern is that inspiral-merger information is inferred under the Kerr assumption, raising questions about its consistency when constraining non-Kerr parameters on the ringdown phase. However, additional hair arising from unknown merger dynamics would not induce significant macroscopic departures from Kerr behavior, thus using inspiral-merger-inferred distributions as the informative priors is a physically well-motivated and reasonable realization.

Future studies should focus on extending the present framework to the case of $s=-2$ perturbation and performing hierarchical Bayesian analyses. Our findings highlight that quantum corrections, though small at the level of the background metric, may have non-negligible effects on ringdown-based parameter inference. This opens a promising avenue for testing quantum-gravity-induced deviations using gravitational-wave spectroscopy~\cite{Berti:2025hly}. With the advent of next-generation detectors such as the Einstein Telescope~\cite{ET:2019dnz}, Cosmic Explorer~\cite{Reitze:2019iox}, LISA~\cite{LISA:2017pwj}, and Taiji~\cite{Hu:2017mde}, the precision of parameter measurements of black holes will increase significantly, making it possible to probe the quantum parameter, such as $\bar{\alpha}$ in the RQCBH mode, through multi-mode or multi-event analyses.

\appendix
\section{Hyperboloidal coordinates of the rotating quantum corrected black holes}\label{height_functions}
In this appendix, following~\cite{PanossoMacedo:2019npm}, we construct hyperboloidal coordinates of the rotating quantum corrected black holes. We start from the ingoing Kerr coordinates $\{v,r,\theta,\phi\}$ via
\begin{eqnarray}\label{ingoing_Kerr_coordinates}
	t=v-r_{\star}\, ,\quad \varphi=\phi-k(r)\, ,
\end{eqnarray}
with the tortoise $r_{\star}(r)$ and the phase $k(r)$ defined by
\begin{eqnarray}
	\frac{\mathrm{d}r_{\star}}{\mathrm{d}r}=\frac{r^2+a^2}{\Delta}\, ,\quad \frac{\mathrm{d}k}{\mathrm{d}r}=\frac{a}{\Delta}\, .
\end{eqnarray}
Therefore, the metric (\ref{BL_coordinate}) in terms of the ingoing Kerr coordinates is transformed into
\begin{eqnarray}
	\mathrm{d}s^2&=&-\Big(\frac{\Delta-a^2\sin^2\theta}{\Sigma}\Big)\Big(\mathrm{d}v-\frac{r^2+a^2}{\Delta}\mathrm{d}r\Big)^2\nonumber\\
    &&-2a\sin^2\theta\Big(1-\frac{\Delta-a^2\sin^2\theta}{\Sigma}\Big)\Big(\mathrm{d}v-\frac{r^2+a^2}{\Delta}\mathrm{d}r\Big)\Big(\mathrm{d}\phi-\frac{a}{\Delta}\mathrm{d}r\Big)\nonumber\\
	&&+\sin^2\theta\Big[\Sigma+a^2\sin^2\theta\Big(2-\frac{\Delta-a^2\sin^2\theta}{\Sigma}\Big)\Big]\Big(\mathrm{d}\phi-\frac{a}{\Delta}\mathrm{d}r\Big)^2+\frac{\Sigma}{\Delta}\mathrm{d}r^2+\Sigma\mathrm{d}\theta^2\nonumber\\
	&=&-\Big(\frac{\Delta-a^2\sin^2\theta}{\Sigma}\Big)\Big(\mathrm{d}v-a\sin^2\theta\mathrm{d}\phi\Big)^2+\Sigma\Big(\mathrm{d}\theta^2+\sin^2\theta\mathrm{d}\phi^2\Big)\nonumber\\
    &&+2\Big(\mathrm{d}v-a\sin^2\theta\mathrm{d}\phi\Big)\Big(\mathrm{d}r-a\sin^2\theta\mathrm{d}\phi\Big)\, .
\end{eqnarray}
By the intrinsic spacetime geometry, the hypersurface defined by $r = r_{+}$ at fixed $v$ constitutes the future black-hole horizon, whereas the asymptotic limit $r\to\infty$ converges to past null infinity  $\mathscr{I}^{-}$.

For the characterization of $\mathscr{I}^{+}$,  monitoring the null vectors $k^a$ and $l^a$ corresponding to ingoing and outgoing light rays respectively proves advantageous. In the extant coordinate system $\{v, r, \theta, \phi\}$, these vectors admit the explicit representation:
\begin{eqnarray}\label{null_vector_ingoing}
	k^\mu=\zeta(0,-1,0,0)\, ,\quad l^\mu=\zeta^{-1}\Big(\frac{r^2+a^2}{\Sigma},\frac{\Delta}{2\Sigma},0,\frac{a}{\Sigma}\Big)\, ,
\end{eqnarray}
where the parameter $\zeta$ is the boost parameter which will be determined in the followings. In fact, it's not difficult to notice that two null vectors satisfy
\begin{eqnarray}
	g_{ab}k^ak^b=g_{ab}l^al^b=0\, ,\quad g_{ab}k^al^b=-1\, .
\end{eqnarray}

Finally, we introduce compact hyperboloidal coordinates $\{\tau,\sigma,\theta,\phi\}$ via the height function technique~\cite{PanossoMacedo:2019npm}, which is given by
\begin{eqnarray}
	v(\tau,\sigma,\theta)=\lambda\Big[\tau-h(\sigma,\theta)\Big]\, ,\quad r(\sigma)=\lambda\frac{\rho(\sigma)}{\sigma}\, .
\end{eqnarray}
The second equation is the radial compactification, which allows us to naturally associate a conformal factor in terms of the new coordinate $\sigma$ via
\begin{eqnarray}
	\Omega=\sigma/\lambda\, .
\end{eqnarray}

In the conformal spacetime, the conformal null vectors are rescaled into $\tilde{k}^a=\Omega^{-1}k^a$ and $\tilde{l}^a=\Omega^{-1}l^a$. To ensure that the hypersurfaces $\tau=\text{constant}$ foliate future null infinity, it is required that $\tau$ is a good parameter of the ingoing conformal null vector via $\tilde{k}^a\partial_a\tau=\tilde{k}^{\tau}=1$. This requirement fixes the boost parameter $\zeta$ of Eqs. (\ref{null_vector_ingoing}), and it leads to
\begin{eqnarray}
	\tilde{k}^\tau=\Omega^{-1}k^{\tau}=\Omega^{-1}\frac{\partial\tau}{\partial r}k^r=\frac{\sigma h_{,\sigma}}{\beta}\zeta=1\, ,\quad \tilde{k}^{\sigma}=\Omega^{-1}k^{\sigma}=\frac{\partial\sigma}{\partial r}k^r=\frac{\sigma}{\beta}\zeta=\frac{1}{h_{,\sigma}}\, ,
\end{eqnarray}
where we use the boost parameter $\zeta$ given by
\begin{eqnarray}
	\zeta=\frac{\beta}{\sigma h_{,\sigma}}\, ,\quad \beta(\sigma)=\rho(\sigma)-\sigma\rho^{\prime}(\sigma)\, .
\end{eqnarray}
At the same time, the conformal vector $\tilde{l}^a$ is given by
\begin{eqnarray}
	\tilde{l}^\tau&=&\Omega^{-1}l^\tau=\Omega^{-1}\Big(\frac{\partial\tau}{\partial v}l^v+\frac{\partial\tau}{\partial r}l^r\Big)=\frac{\lambda}{\sigma}\zeta^{-1}\Big(\frac{r^2+a^2}{\lambda\Sigma}-\frac{\sigma^2h_{.\sigma}}{\lambda\beta}\frac{\Delta}{2\Sigma}\Big)=\frac{h_{,\sigma}}{2\beta^2\Sigma}\Big[2\beta(r^2+a^2)-\Delta\sigma^2h_{,\sigma}\Big]\, ,\label{tilde_l_tau}\\
	\tilde{l}^{\sigma}&=&\Omega^{-1}l^\sigma=\Omega^{-1}\frac{\partial\sigma}{\partial r}l^r=\frac{\lambda}{\sigma}\Big(-\frac{\sigma^2}{\lambda\beta}\Big)\zeta^{-1}\frac{\Delta}{2\Sigma}=-\frac{\Delta\sigma^2h_{,\sigma}}{2\beta^2\Sigma}\, ,\\
	\tilde{l}^{\phi}&=&\Omega^{-1}l^{\phi}=\frac{\lambda}{\sigma}\zeta^{-1}\frac{a}{\Sigma}=\frac{\lambda ah_{,\sigma}}{\beta\Sigma}\, .
\end{eqnarray}
Finally, we impose that $\sigma= 0$ is a null surface corresponding to future null infinity via
\begin{eqnarray}
	\lim_{\sigma\to0}\frac{1}{h_{,\sigma}}=0\, .
\end{eqnarray}
Nevertheless, the stipulated condition must preserve  the regularity of the outgoing conformal null generator $\tilde{l}^a$ in the limit $\sigma \to 0$, which is an important condition to determine the expression of the height function. It is natural to consider the function $\rho(\sigma)$ to be a regular function, i.e., around $\sigma=0$, the function $\rho(\sigma)$ satisfies
\begin{eqnarray}\label{function_rho}
	\rho(\sigma)=\rho_0+\rho_1\sigma+\rho_2\sigma^2+\mathcal{O}(\sigma^3)\, ,
\end{eqnarray}
and then the function $\beta(\sigma)$ satisfies
\begin{eqnarray}\label{function_beta}
	\beta(\sigma)=\rho_0-\rho_2\sigma^2+\mathcal{O}(\sigma^3)\, .
\end{eqnarray}
Substituting Eq. (\ref{function_rho}) and Eq. (\ref{function_beta}) into Eq. (\ref{tilde_l_tau})   and remaining the component $\tilde{l}^{\tau}$ finite, one gets
\begin{eqnarray}
	h_{,\sigma}=\frac{2\rho_0}{\sigma^2}+\frac{4M}{\lambda\sigma}+\mathcal{O}(1)\, .
\end{eqnarray}
Then, the height function $h(\sigma,\theta)$ has the general form
\begin{eqnarray}
	h(\sigma,\theta)=-\frac{2\rho_0}{\sigma}+\frac{4M}{\lambda}\ln\sigma+A(\sigma,\theta)\, .
\end{eqnarray}

\section{The expressions of the operator $L_1$ and $L_2$}\label{expressions_L1_and_L2}
In this appendix, we give the explicit expression of the operator $L$ in Eq. (\ref{operator_L}). The expressions of the operators $L_1$ and  $L_2$ are written as
\begin{eqnarray}
	L_1={}^{[x]}L_1^{2}(\sigma,x)\frac{\partial^2}{\partial x^2}+{}^{[\sigma]}L_1^{2}(\sigma,x)\frac{\partial^2}{\partial \sigma^2}+{}^{[x]}L_1^{1}(\sigma,x)\frac{\partial}{\partial x}+{}^{[\sigma]}L_1^{1}(\sigma,x)\frac{\partial}{\partial \sigma}+L_1^0(\sigma,x)\, ,
\end{eqnarray}
\begin{eqnarray}
	L_2={}^{[\sigma]}L_2^{1}(\sigma,x)\frac{\partial}{\partial \sigma}+L_2^0(\sigma,x)\, ,
\end{eqnarray}
where seven functions ${}^{[x]}L_1^{2}(\sigma,x)$, ${}^{[\sigma]}L_1^{2}(\sigma,x)$, ${}^{[x]}L_1^{1}(\sigma,x)$, ${}^{[\sigma]}L_1^{1}(\sigma,x)$, $L_1^0(\sigma,x)$, ${}^{[\sigma]}L_2^{1}(\sigma,x)$, and $L_2^0(\sigma,x)$ associated with $(\sigma,x)$ are
\begin{eqnarray}
	{}^{[x]}L_1^{2}(\sigma,x)\nonumber&=&\Bigg\{\kappa^2(x^2-1)-\frac{4}{(q^4+q^2+1)^3}\Big[\kappa ^2 \sigma +q^6 \sigma +q^4 (\sigma +1)+q^2(\kappa ^2 \sigma +\sigma +1)+\sigma +1\Big]\nonumber\\
    &&\times\Bigg[(\kappa ^2+1) \Big(\kappa ^2 (\sigma -1)-1\Big)+q^{12}(\sigma ^3-1)+q^{10}\Big(\kappa^2(\sigma -\sigma ^3)+\sigma ^3+\sigma ^2-2\Big)\nonumber\\
    &&+q^8 \Big(-\kappa ^2(\sigma^2-2\sigma +2)+\sigma ^3+\sigma ^2-3\Big)+q^6\Big(\kappa ^4(\sigma-\sigma^3)-\kappa^2(\sigma ^2-3 \sigma +4)+\sigma ^3+\sigma ^2-4\Big)\nonumber\\
    &&-q^4 \Big(\kappa ^4(\sigma ^3-2 \sigma +1)+\kappa ^2(\sigma ^3+\sigma ^2-3 \sigma +4)+3\Big)\nonumber\\
    &&+2(\kappa ^2+1) q^2 \Big(\kappa ^2 (\sigma -1)-1\Big)\Bigg]\Bigg\}^{-1}\times(1-x^2)\, ,
\end{eqnarray}
\begin{eqnarray}
	{}^{[\sigma]}L_1^{2}(\sigma,x)&=&\Bigg\{(q^4+q^2+1)\Bigg[\kappa^2(x^2-1)-\frac{4}{(q^4+q^2+1)^3}\Big[\kappa^2\sigma+q^6\sigma +q^4(\sigma+1)+q^2(\kappa^2\sigma+\sigma +1)+\sigma+1\Big]\nonumber\\
    &&\times\Big[(\kappa ^2+1)\Big(\kappa ^2 (\sigma -1)-1\Big)+q^{12}(\sigma ^3-1)+q^{10} \Big(\kappa ^2(\sigma-\sigma^3)+\sigma ^3+\sigma ^2-2\Big)\nonumber\\
    &&+q^8\Big(-\kappa^2(\sigma ^2-2 \sigma +2)+\sigma ^3+\sigma ^2-3\Big)+q^6 \Big(\kappa ^4 (\sigma-\sigma ^3)-\kappa^2(\sigma ^2-3 \sigma +4)+\sigma ^3+\sigma ^2-4\Big)\nonumber\\
    &&-q^4 \Big(\kappa ^4(\sigma ^3-2 \sigma +1)+\kappa ^2(\sigma ^3+\sigma ^2-3 \sigma +4)+3\Big)+2(\kappa ^2+1) q^2 \Big(\kappa ^2 (\sigma -1)-1\Big)\Big]\Bigg]\Bigg\}^{-1}\nonumber\\
    &&\times\Big[(\sigma -1)\sigma^2(q^2 \sigma -1) \Big(-\kappa^2\sigma +q^4(\sigma ^2+\sigma +1)-q^2 (\sigma +1)(\kappa^2 \sigma -1)+1\Big)\Big]\, ,
\end{eqnarray}
\begin{eqnarray}
	{}^{[x]}L_1^{1}(\sigma,x)&=&-\Bigg\{\kappa^2(x^2-1)-\frac{4}{(q^4+q^2+1)^3}\Big[\kappa^2\sigma +q^6\sigma+q^4(\sigma+1)+q^2(\kappa^2\sigma +\sigma+1)+\sigma +1\Big]\nonumber\\
    &&\times\Big[(\kappa^2+1) \Big(\kappa ^2 (\sigma -1)-1\Big)+q^{12}(\sigma^3-1)+q^{10} \Big(\kappa^2(\sigma-\sigma^3)+\sigma^3+\sigma ^2-2\Big)\nonumber\\
    &&+q^8\Big(-\kappa^2(\sigma^2-2\sigma +2)+\sigma^3+\sigma^2-3\Big)+q^6 \Big(\kappa^4(\sigma -\sigma ^3)-\kappa^2(\sigma ^2-3 \sigma +4)+\sigma ^3+\sigma ^2-4\Big)\nonumber\\
    &&-q^4\Big(\kappa ^4(\sigma ^3-2 \sigma +1)+\kappa ^2(\sigma ^3+\sigma ^2-3 \sigma +4)+3\Big)+2(\kappa ^2+1) q^2 \Big(\kappa ^2 (\sigma -1)-1\Big)\Big]\Bigg\}^{-1}\nonumber\\
    &&\times\Big[2x(|m|+1)\Big]\, ,
\end{eqnarray}
\begin{eqnarray}
	{}^{[\sigma]}L_1^{1}(\sigma,x)&=&-\sigma\Bigg\{(q^4+q^2+1)\Bigg[\kappa ^2(x^2-1)-\frac{4}{(q^4+q^2+1)^3}\Big[\kappa^2\sigma+q^6 \sigma +q^4 (\sigma+1)+q^2(\kappa^2\sigma+\sigma +1)+\sigma +1\Big]\nonumber\\
    &&\times\Big[(\kappa ^2+1) \Big(\kappa ^2 (\sigma -1)-1\Big)+q^{12}(\sigma ^3-1)+q^{10} \Big(\kappa ^2(\sigma-\sigma^3)+\sigma ^3+\sigma ^2-2\Big)\nonumber\\
    &&+q^8 \Big(-\kappa^2(\sigma ^2-2\sigma +2)+\sigma ^3+\sigma ^2-3\Big)+q^6 \Big(\kappa ^4(\sigma -\sigma ^3)-\kappa ^2(\sigma ^2-3 \sigma +4)+\sigma ^3+\sigma ^2-4\Big)\nonumber\\
    &&-q^4 \Big(\kappa ^4(\sigma ^3-2 \sigma +1)+\kappa ^2(\sigma ^3+\sigma ^2-3 \sigma +4)+3\Big)+2(\kappa ^2+1) q^2 \Big(\kappa ^2 (\sigma -1)-1\Big)\Big]\Bigg]\Bigg\}^{-1}\nonumber\\
    &&\times\Big[-4\kappa^2\sigma ^2+\sigma(3 \kappa^2+2\mathrm{i}\kappa m+3)+q^4 \Big(6 \kappa ^2 \sigma ^4-4 \kappa ^2 \sigma ^2+\sigma  (3+2\mathrm{i}\kappa m)-2\Big)\nonumber\\
    &&+q^2 \Big(-4 \kappa ^2 \sigma ^2+\sigma(3 \kappa ^2+2 \mathrm{i} \kappa  m+3)-2\Big)+q^6(3 \sigma -6 \sigma ^4)-2\Big]\, ,
\end{eqnarray}
\begin{eqnarray}
	L_1^0(\sigma,x)&=&-\Bigg\{\kappa ^2(x^2-1)-\frac{4}{(q^4+q^2+1)^3}\Big[\kappa ^2 \sigma+q^6\sigma+q^4(\sigma +1)+q^2(\kappa^2\sigma +\sigma +1)+\sigma+1\Big]\nonumber\\
    &&\times\Bigg[(\kappa ^2+1) \Big(\kappa^2(\sigma-1)-1\Big)+q^{12}(\sigma ^3-1)+q^{10}\Big(\kappa ^2(\sigma -\sigma ^3)+\sigma ^3+\sigma ^2-2\Big)\nonumber\\
    &&+q^8 \Big(-\kappa ^2(\sigma ^2-2 \sigma +2)+\sigma ^3+\sigma ^2-3\Big)+q^6 \Big(\kappa ^4(\sigma -\sigma ^3)-\kappa ^2(\sigma ^2-3 \sigma +4)+\sigma ^3+\sigma ^2-4\Big)\nonumber\\
    &&-q^4\Big(\kappa ^4(\sigma ^3-2 \sigma +1)+\kappa ^2(\sigma ^3+\sigma ^2-3 \sigma +4)+3\Big)+2(\kappa ^2+1) q^2 \Big(\kappa^2(\sigma -1)-1\Big)\Bigg]\Bigg\}^{-1}\nonumber\\
    &&\times\Bigg[|m|^2+|m|+\frac{\sigma}{q^4+q^2+1}\Big[-2\kappa^2\sigma+\kappa ^2+2\mathrm{i}\kappa m+q^4\Big(\kappa^2(4\sigma ^3-2 \sigma)+2\mathrm{i}\kappa  m+1\Big)\nonumber\\
    &&+q^2\Big(\kappa^2(1-2\sigma)+2\mathrm{i} \kappa m+1\Big)+q^6(1-4 \sigma ^3)+1\Big]\Bigg]\, ,
\end{eqnarray}
\begin{eqnarray}
	{}^{[\sigma]}L_2^{1}(\sigma,x)&=&-\Bigg\{\kappa ^2(x^2-1)-\frac{4}{\left(q^4+q^2+1\right)^3}\Big[\kappa ^2 \sigma+q^6\sigma+q^4(\sigma+1)+q^2(\kappa ^2 \sigma +\sigma+1)+\sigma+1\Big]\nonumber\\
    &&\times\Bigg[(\kappa ^2+1) \Big(\kappa ^2 (\sigma -1)-1\Big)+q^{12} (\sigma ^3-1)+q^{10} \Big(\kappa ^2 \left(\sigma -\sigma ^3\right)+\sigma ^3+\sigma ^2-2\Big)\nonumber\\
    &&+q^8 \Big(-\kappa ^2(\sigma ^2-2 \sigma +2)+\sigma ^3+\sigma ^2-3\Big)+q^6 \Big(\kappa^4(\sigma -\sigma ^3)-\kappa^2(\sigma ^2-3 \sigma +4)+\sigma ^3+\sigma ^2-4\Big)\nonumber\\
    &&-q^4 \Big(\kappa ^4(\sigma ^3-2 \sigma +1)+\kappa ^2(\sigma ^3+\sigma ^2-3 \sigma +4)+3\Big)+2(\kappa ^2+1) q^2 \Big(\kappa ^2 (\sigma -1)-1\Big)\Bigg]\Bigg\}^{-1}\nonumber\\
    &&\times\Bigg\{2(\kappa ^2 \sigma ^2+1)\Bigg[1-\Big[(q^4+q^2+1)^2(\kappa ^2 \sigma ^2+1)\Big]^{-1}\times\Big[2 (\sigma -1)(q^2 \sigma -1) \Big(\kappa ^2 \sigma +q^6 \sigma +q^4 (\sigma +1)\nonumber\\
    &&+q^2(\kappa ^2 \sigma +\sigma +1)+\sigma +1\Big) \Big(-\kappa ^2 \sigma +q^4(\sigma ^2+\sigma +1)-q^2 (\sigma +1)(\kappa ^2 \sigma -1)+1\Big)\Big]\Bigg]\Bigg\}\, ,
\end{eqnarray}
\begin{eqnarray}
	L_2^0(\sigma,x)&=&-\Bigg\{(q^4+q^2+1)^2 \Bigg[\kappa^2(x^2-1)-\frac{4}{(q^4+q^2+1)^3}\Big[\kappa^2\sigma +q^6 \sigma +q^4 (\sigma +1)+q^2(\kappa^2\sigma +\sigma +1)+\sigma +1\Big]\nonumber\\
    &&\times\Big[(\kappa ^2+1) \Big(\kappa ^2 (\sigma -1)-1\Big)+q^{12}(\sigma ^3-1)+q^{10} \Big(\kappa ^2 \left(\sigma -\sigma ^3\right)+\sigma ^3+\sigma ^2-2\Big)\nonumber\\
    &&+q^8 \Big(-\kappa ^2(\sigma ^2-2 \sigma +2)+\sigma ^3+\sigma ^2-3\Big)+q^6 \Big(\kappa ^4(\sigma -\sigma ^3)-\kappa ^2(\sigma ^2-3 \sigma +4)+\sigma ^3+\sigma ^2-4\Big)\nonumber\\
    &&-q^4 \Big(\kappa^4(\sigma^3-2\sigma +1)+\kappa ^2(\sigma ^3+\sigma^2-3\sigma +4)+3\Big)+2(\kappa ^2+1) q^2 \Big(\kappa^2(\sigma -1)-1\Big)\Big]\Bigg]\Bigg\}^{-1}\nonumber\\
    &&\times\Bigg\{2\sigma\Bigg[\kappa^4(2-3 \sigma )-3 \kappa ^2 (\sigma -1)+q^{12}(2-5 \sigma ^3)+q^{10} \Big(\kappa ^2(5 \sigma ^2-3) \sigma -5 \sigma ^3-4 \sigma ^2+4\Big)\nonumber\\
    &&+q^8 \Big(\kappa ^2(4 \sigma ^2-6 \sigma +3)-5 \sigma ^3-4 \sigma ^2+6\Big)+q^6 \Big(\kappa ^4 \sigma(5 \sigma ^2-3)+\kappa ^2(4 \sigma ^2-9 \sigma +6)-5 \sigma ^3-4 \sigma ^2+8\Big)\nonumber\\
    &&+q^4 \Big(\kappa ^4(5 \sigma ^3-6 \sigma +2)+\kappa ^2(5 \sigma ^3+4 \sigma ^2-9 \sigma +5)+6\Big)+q^2 \Big(\kappa ^4 (4-6 \sigma )-6 \kappa ^2 (\sigma -1)+4\Big)+2\Bigg]\nonumber\\
    &&+2 \mathrm{i} \kappa  m (q^4+q^2+1) \Big[2(\kappa ^2+1) \sigma +2 q^6 \sigma +q^4 (2 \sigma +1)+q^2 \Big(2(\kappa ^2+1) \sigma +1\Big)+1\Big]\Bigg\}\, .
\end{eqnarray}

\section{The numerical accuracy test}\label{numerical_accuracy_test}
In this appendix, we will give the numerical accuracies for the QNM spectra fittings and the convergence test of the QNMs. As for the convergence test for the choice of resolution in the pseudo-spectral method, we use three resolutions namely (1) $N_\sigma=20$, $N_x=10$, (2) $N_\sigma=18$, $N_x=10$, (3) $N_\sigma=20$, $N_x=8$ to show the test results. The relative error of different resolutions is defined as
\begin{eqnarray}\label{relative_error}
    \delta\omega=\frac{|\omega_{\text{test}}-\omega_{\text{benchmark}}|}{|\omega_{\text{benchmark}}|}\, ,
\end{eqnarray}
where we choose $N_\sigma=20$, $N_x=10$ as the benchmark resolution. Two groups of parameters $(\kappa,q)$ are used to display the results of $\delta\omega$, which are shown in Tab. \ref{convergence_test_QNMs}.
\begin{table}[ht]
 \renewcommand{\arraystretch}{1.5}
 \centering
\begin{tabular}{cccccc}
 \hline
   \hline
~~~~~~&\multicolumn{2}{c}{$\kappa=0.2$\quad $q=0.4$} ~~~~~&\multicolumn{2}{c}{$\kappa=0.4$\quad $q=0.8$}~~~~~\\
\cmidrule(l{2pt}r{2pt}){2-3}\cmidrule(l{2pt}r{2pt}){4-5}
~~~~~& $\delta\omega (N_\sigma=18, N_x=10)$ ~~~~~& $\delta\omega (N_\sigma=20, N_x=8)$~~~~~&$\delta\omega (N_\sigma=18, N_x=10)$~~~~~& $\delta\omega (N_\sigma=20, N_x=8)$\\
 \hline
$\omega_{220}^{+}$  ~~~~&$3.105\times10^{-8}$~~~~&$2.200\times10^{-20}$~~~~&$1.425\times10^{-7}$~~~~&$2.203\times10^{-18}$& \\
$\omega_{220}^{-}$ ~~~~& $8.519\times10^{-6}$ ~~~~&$7.281\times10^{-20}$~~~~&$2.399\times10^{-5}$~~~~&$7.663\times10^{-18}$&\\
$\omega_{221}^{+}$  ~~~~&$2.881\times10^{-9}$~~~~&$2.518\times10^{-15}$~~~~&$1.513\times10^{-8}$~~~~&$1.175\times10^{-13}$& \\
$\omega_{221}^{-}$ ~~~~& $5.214\times10^{-7}$ ~~~~&$4.035\times10^{-15}$~~~~&$1.480\times10^{-6}$~~~~&$1.869\times10^{-13}$&\\
$\omega_{320}^{+}$  ~~~~&$4.477\times10^{-9}$~~~~&$1.827\times10^{-19}$~~~~&$2.160\times10^{-9}$~~~~&$1.207\times10^{-16}$& \\
$\omega_{320}^{-}$ ~~~~& $1.024\times10^{-6}$ ~~~~&$3.959\times10^{-19}$~~~~&$1.687\times10^{-7}$~~~~&$1.573\times10^{-16}$&\\
$\omega_{321}^{+}$  ~~~~&$6.999\times10^{-10}$~~~~&$8.299\times10^{-15}$~~~~&$7.201\times10^{-10}$~~~~&$1.101\times10^{-12}$& \\
$\omega_{321}^{-}$ ~~~~& $1.124\times10^{-7}$ ~~~~&$1.169\times10^{-14}$~~~~&$4.323\times10^{-8}$~~~~&$1.286\times10^{-12}$&\\
   \hline
 \hline
\end{tabular}
\caption{The relative errors of QNM spectra between different resolutions, where $\kappa=0.2$, $q=0.4$ and $\kappa=0.4$ and $q=0.8$ are used.}
\label{convergence_test_QNMs}
\end{table}

As for the fitting accuracies of the QNMs, we give the maximum residuals of the QNM spectra fittings from \textit{Mathematica} by using the built-in function \textit{NonlinearModelFit}. The results are shown in the Tab. \ref{maximum_residual_spectra_fitting}, where $N_1=4$ and $N_2=4$ are taken in the model (\ref{fitting_model}).  

\begin{table}[ht]
 \renewcommand{\arraystretch}{1.5}
 \centering
\begin{tabular}{lccccc}
 \hline
   \hline
~~~~~~& Maximum residuals ~~~~~~& \cr
 \hline
$\text{Re}\omega_{220}^{+}$  ~~~~~~~~~~~~&$8.497\times10^{-8}$& \\
$\text{Im}\omega_{220}^{+}$  ~~~~~~~~~~~~& $2.402\times10^{-7}$& \\
$\text{Re}\omega_{220}^{-}$ ~~~~~~~~~~~~& $2.304\times10^{-5}$ &\\
$\text{Im}\omega_{220}^{-}$ ~~~~~~~~~~~~& $1.419\times10^{-5}$ &\\
\hline
$\text{Re}\omega_{221}^{+}$  ~~~~~~~~~~~~&$7.414\times10^{-6}$& \\
$\text{Im}\omega_{221}^{+}$  ~~~~~~~~~~~~& $3.598\times10^{-6}$& \\
$\text{Re}\omega_{221}^{-}$ ~~~~~~~~~~~~& $1.178\times10^{-4}$ &\\
$\text{Im}\omega_{221}^{-}$ ~~~~~~~~~~~~& $1.473\times10^{-4}$ &\\
   \hline
$\text{Re}\omega_{320}^{+}$  ~~~~~~~~~~~~&$3.347\times10^{-8}$& \\
$\text{Im}\omega_{320}^{+}$  ~~~~~~~~~~~~& $2.104\times10^{-7}$& \\
$\text{Re}\omega_{320}^{-}$ ~~~~~~~~~~~~& $5.642\times10^{-6}$ &\\
$\text{Im}\omega_{320}^{-}$ ~~~~~~~~~~~~& $5.027\times10^{-6}$ &\\
\hline
$\text{Re}\omega_{321}^{+}$  ~~~~~~~~~~~~&$7.815\times10^{-7}$& \\
$\text{Im}\omega_{321}^{+}$  ~~~~~~~~~~~~& $5.296\times10^{-7}$& \\
$\text{Re}\omega_{321}^{-}$ ~~~~~~~~~~~~& $6.500\times10^{-5}$ &\\
$\text{Im}\omega_{321}^{-}$ ~~~~~~~~~~~~& $1.223\times10^{-4}$ &\\
   \hline
 \hline
\end{tabular}
\caption{The maximum residuals of the QNM spectra fittings for the model (\ref{fitting_model}) with $N_1=4$ and $N_2=4$.}
\label{maximum_residual_spectra_fitting}
\end{table}

\section*{Acknowledgement}
The code used for parameter estimation using \texttt{pyRing} in this work is available online~\cite{Chen2025_pyRing_RQCBH}. This work is supported in part by the National Key R\&D Program of China Grant No. 2022YFC2204603. It is also supported by the National Natural Science Foundation of China with grants No. 12475063, No. 12075232 and No. 12247103. This work is also supported by the National Natural Science Foundation of China with grant No. 12505067. The Python package \texttt{pyRing} relies on other open-source packages \texttt{corner,cpnest,gwpy,lalsuite,matplotib,numpy,scipy,} and \texttt{pesummary}~\cite{corner, john_veitch_2025_15504752, MACLEOD2021100657, swiglal, lalsuite, Hunter:2007, harris2020array, 2020SciPy-NMeth, Hoy:2020vys}.

\bibliography{reference}
\bibliographystyle{apsrev4-1}

\end{document}